%% file: manuscript.tex
\documentclass[aps,prc,twocolumn,nofootinbib]{revtex4-1}
\usepackage[utf8]{inputenc}
\usepackage{csquotes}
\usepackage{todonotes}
\usepackage{amsmath,amssymb}
\usepackage{physics}
\usepackage{currfile}
\usepackage{bbold}
\usepackage{textcomp}

\usepackage{hyperref}

\input{in/macros}

\begin{document}
\title{Bayesian parameter estimation in $\chi$EFT using Hamiltonian Monte Carlo}
\date{\today}
\author{Isak Svensson}
\email{isak.svensson@chalmers.se}
\affiliation{Department of Physics, Chalmers University of Technology, SE-412 96 G\"oteborg, Sweden}

\author{Andreas Ekstr\"om}
\affiliation{Department of Physics, Chalmers University of Technology, SE-412 96 G\"oteborg, Sweden}

\author{Christian Forss\'en}
\affiliation{Department of Physics, Chalmers University of Technology, SE-412 96 G\"oteborg, Sweden}

\begin{abstract}
The number of low-energy constants (LECs) in chiral effective field theory (\chieft) grows rapidly with increasing chiral order, necessitating the use of Markov chain Monte Carlo techniques for sampling their posterior probability density function. For this we introduce a Hamiltonian Monte Carlo (HMC) algorithm and sample the LEC posterior up to next-to-next-to-leading order (\nnlo) in the two-nucleon sector of \chieft. We find that the sampling efficiency of HMC is three to six times higher compared to an affine-invariant sampling algorithm. We analyze the empirical coverage probability and validate that the \nnlo{} model yields predictions for two-nucleon scattering data with largely reliable credible intervals, provided that one ignores the leading order EFT expansion parameter when inferring the variance of the truncation error. We also find that the \nnlo{} truncation error dominates the error budget.
\end{abstract}

\maketitle

\section{Introduction}
Chiral effective field theory (\chieft) descriptions~\cite{Bedaque:2002mn, Epelbaum:2008ga, Machleidt:2011zz, Hammer:2019poc} of the strong nuclear interaction depend on low-energy constants (LECs) that govern the strength of the various interaction terms. Their numerical values must be inferred from data and are best described by a posterior probability density function (\pdf). Clearly, this parametric uncertainty will combine with the inherent discrepancy of \chieft, i.e., the epistemic gap between model predictions and real world observations. Thus, their joint uncertainty must be estimated and propagated forward before the credibility of the underlying theory and its predictions for nuclear observables can be assessed. Drawing samples from this posterior predictive distribution (\ppd) is key to analyze the implications of physical and probabilistic modeling choices in the \emph{ab initio} description of nuclear systems. Fortunately, operating with a \chieft{} endowed with a power counting offers a principal handle to estimate the relevant model discrepancy in terms of Bayesian credible intervals for the EFT truncation error~\cite{furnstahl15}. Still, drawing samples from relevant posterior \pdf s presents a formidable challenge, particularly for high-dimensional parameter volumes.

Here, we introduce the Hamiltonian Monte Carlo (HMC) method~\cite{duane87} to sample LEC posteriors and \ppd s in \chieft. HMC is a Markov chain Monte Carlo (MCMC)~\cite{metropolis53,Hastings:1970aa} method that exploits the equations of Hamiltonian dynamics for drawing uncorrelated \pdf{} samples with a high acceptance probability. Crucially, HMC performs well also in cases of high-dimensional probability distributions where most other MCMC algorithms fail to converge within a reasonable time frame. In this paper we demonstrate how to use HMC to efficiently sample the LEC posteriors at leading order (\lo), next-to-leading order (\nlo), and next-to-next-to-leading order (\nnlo) in deltaless \chieft{}. We also perform model checking and model validation by drawing samples from the \ppd{} for elastic nucleon-nucleon (\NN) scattering in the neutron-proton (\np) and proton-proton (\pp) channels. All \pdf s are conditioned on data from the recent Granada database~\cite{perez13-1, perez13-2} of measured \NN{} scattering cross sections. We employ a statistical model for the EFT truncation error, proposed by~\citet{wesolowski19}, to account for the model discrepancy due to excluded contributions from higher chiral orders.

In a Bayesian data analysis it is straightforward to condition on existing results. We make use of a previous Roy-Steiner analysis~\cite{Hoferichter:2015hva,siemens17} to incorporate prior knowledge about the LECs that govern the strengths of subleading pion-nucleon (\piN{}) interactions. We also place a prior in accordance with naturalness expectations on the LECs that govern contact interaction strengths. Throughout this work, we define the necessary one- and two-pion exchanges and contact interactions according to Ref.~\cite{Machleidt:2011zz} and follow the same conventions for the potential, scattering amplitudes, and \NN{} scattering observables as in Ref.~\cite{Carlsson:2015vda}. We use a non-local super-Gaussian momentum-space regulator with a fixed cutoff $\Lambda=450$ MeV.


\section{Bayesian parameter estimation in $\chi$EFT \label{sec:bayesian_estimation}}
Bayes' theorem
\begin{equation}
\label{eq:bayes}
\prob(\lecs|\Dtrain,I) = \frac{\prob(\Dtrain|\lecs,I) \cdot \prob(\lecs|I)}{\prob(\Dtrain|I)},
\end{equation}
provides a straightforward way to express the posterior \pdf{} $\prob(\lecs | \Dtrain, I)$ for the LECs $\lecs$ in terms of a likelihood $\prob(\Dtrain|\lecs,I)$, prior $\prob(\lecs|I)$, and marginal likelihood (or evidence) $\prob(\Dtrain|I)$. In this work, the LEC posterior is conditioned on \NN{} scattering data $\Dtrain$ and additional prior information $I$. The marginal likelihood, which does not depend on $\lecs$, plays no role in parameter estimation and we therefore have
\begin{equation}
\label{eq:bayes2}
\prob(\lecs|\Dtrain,I) \propto \prob(\Dtrain|\lecs,I) \cdot \prob(\lecs|I).
\end{equation}
A hallmark of the Bayesian approach is the transparent and straightforward inclusion of prior information $I$. A main goal of this work is to incorporate the probabilistic model for the EFT truncation error from Ref.~\cite{wesolowski19} and introduce an HMC algorithm to efficiently sample the posterior \pdf{} for all LECs up to \nnlo{}.

In the following subsections we present the \NN{} data we use, our prior and likelihood, and specify the hyperparameters and details of the model for relating experimental \NN{} scattering data to a \chieft\ prediction at a given chiral order.


\subsection{Prior}
Our full prior for the LECs is written as a product of two independent priors: $\prob(\lecs_{NN} | I)$ for the contact LECs and $\prob(\lecs_{\pi N} | I)$ for the \piN{} LECs. This form,
\begin{equation}
\prob(\lecs | I) = \prob(\lecs_{NN} | I) \cdot \prob(\lecs_{\pi N} | I),
\end{equation}
implies no prior assumption of correlation between LECs from these sectors. Furthermore, we adopt independent and identical Gaussian \pdf{}s for all contact LECs with zero-mean and standard deviation $\bar{\alpha} = 5$. This is a rather weak prior which again makes no assumption of correlations. However, it mildly encodes the naturalness expectation of the LECs by penalizing LEC values $\gg 1$, thus safeguarding somewhat against overfitting~\cite{wesolowski16}. The exact value of $\bar{\alpha}$ does not have a major impact on the outcome, as the large \NN{} data set used in the likelihood strongly dominates over the prior. The prior at \lo\ and \nlo\ in \chieft---where no \piN\ LECs appear---is thus given by
\begin{equation}
\prob(\lecs|I) = \mathcal{N}\left(\vec{0}, \Sigma_\text{prior}\right)
\end{equation}
with
\begin{equation}
\left(\Sigma_\text{prior}\right)_{ij} = \bar{\alpha}^2 \delta_{ij}.
\end{equation}

For the \piN\ LECs $c_1,c_3,c_4$, entering at \nnlo, we have chosen a much more restrictive prior based on mean values and covariance matrices extracted from the maximum-likelihood fit of a Roy-Steiner analysis of the \piN\ scattering amplitudes by~\citet{siemens17}. That analysis proceeds in a kinematical region of chiral perturbation theory that exhibits a stronger curvature with respect to the subthreshold parameters of the \piN\ scattering amplitudes. Once matched to the \piN\ LECs in $\chi$EFT, we obtain a rather informative prior for the corresponding part of the potential. We will return to a more detailed discussion of the \piN\ LECs when analyzing the MCMC posteriors in Sec.~\ref{sec:LEC_posteriors}.

To be specific, the LECs that we consider at each order are:
\begin{equation}
\lecs_\text{LO} = \Bigl(\Ct_{1S0}, \Ct_{3S1}\Bigr)
\end{equation}
\begin{align}
\begin{split}
\lecs_\text{\nlo} &= \Bigl(\Ct_{1S0}^{np}, \Ct_{1S0}^{pp}, \Ct_{3S1}, C_{1S0}, C_{3P0}, \\
&C_{1P1}, C_{3P1}, C_{3S1}, C_{3S1-3D1}, C_{3P2}\Bigr)
\end{split}
\end{align}
\begin{align}
\begin{split}
\lecs_\text{\nnlo} &= \Bigl(c_1, c_3, c_4, \Ct_{1S0}^{np}, \Ct_{1S0}^{pp}, \Ct_{3S1}, C_{1S0}, \\ &C_{3P0}, C_{1P1}, C_{3P1}, C_{3S1}, C_{3S1-3D1}, C_{3P2}\Bigr)
\end{split}
\end{align}
We employ a conventional notation linked to the momentum partial-wave basis, see, e.g., Ref.~\cite{Machleidt:2011zz}. Note also that isospin-breaking effects enter at \nlo{} and only in the $^{1}S_0$ partial-wave.


\subsection{Elastic nucleon-nucleon scattering observables}
\label{sec:data}
We condition our LEC posterior on experimental data. We use roughly two thirds of the Granada 2013 database~\cite{perez13-1,perez13-2} to define the training data set $\Dtrain$. We hold out all scattering data in the range $80 \leq T_\text{lab} \leq 100$ MeV of laboratory scattering energies and assign it to a validation data set $\Dval$. We further assign to $\Dval$ all data in the energy range $290 < T_\text{lab} \leq 350$ MeV, i.e., just above the pion-production threshold, plus the set of integrated \np{} scattering cross sections from Ref.~\cite{lisowski82}. Overall, this choice of data split enables detailed model checking while leaving ample information for estimating the LECs at each chiral order up to \nnlo{}. In all, $\Dtrain$ consists of 4366 experimental data points to be used as input in the parameter estimation process, while $\Dval$ contains 2018 validation data points. The details of the \NN{} scattering data used for parameter estimation and validation are presented in Table~\ref{tab:databases}.

\input{tables/databases}

Given a set of numerical values for the LECs, we compute scattering amplitudes by solving the Lippmann-Schwinger equation for all partial waves with maximum total angular-momentum quantum number $J_\text{max} \leq 30$. This is more than enough to converge the physical model predictions for the resulting scattering observables. Thus, we neglect all sources of numerical or computational method uncertainties going forward.

In \pp{} scattering we include all relevant electromagnetic effects, as outlined in Ref.~\cite{Carlsson:2015vda}: the static Coulomb interaction and its relativistic correction, the first-order approximation to the vacuum polarization, and relevant magnetic moment interactions. This set of long-ranged electromagnetic interaction has been demonstrated by the Nijmegen group to be sufficient for explaining the observed low-energy \pp{} scattering data~\cite{Bergervoet:1988zz,Bergervoet:1990zy}. For this reason, we also neglect any theoretical model discrepancy due to neglected higher-order contributions to the electromagnetic interaction.


\subsection{Likelihood and EFT truncation error}
In this work we relate an experimental measurement $y_\text{exp}$ of some observable $y$ to the true value $y_\text{true}$ via a statistical model
\begin{equation}
y_\text{exp} = y_\text{true} + \delta y_\text{exp}.
\label{eq:yexp}
\end{equation}
This also introduces the experimental uncertainty, $\delta y_\text{exp}$, as a random variable for which we employ the standard deviations provided in the Granada database. We also relate the true value and our theory prediction $y_\text{th}$ via
\begin{equation}
y_\text{true} = y_\text{th} + \delta y_\text{th},
\label{eq:ytrue}
\end{equation}
where $\delta y_\text{th}$ is the model discrepancy term. We will model $\delta y_\text{th}$ as coming from the truncation of the chiral expansion in \chieft{} at some finite chiral order $k$. When doing so we tacitly assume that the entire epistemic uncertainty of the theory can be systematically reduced by going to higher orders in \chieft{}.

To model the truncation error we follow \citet{furnstahl15} and \citet{wesolowski19} and formally write the \chieft{} expansion for some observable prediction $y_\text{th}$ up to chiral order $k$ as
\begin{equation} \label{eq:expansion}
y_\text{th}^{(k)} = y_{\text{ref}} \sum_{\nu = 0}^{k} c_{\nu} Q^\nu,
\end{equation}
where $y_\text{ref}$ is a reference value, $c_{\nu}$ are dimensionless EFT expansion coefficients, and $Q$ is a dimensionless expansion parameter that we assign as
\begin{equation}
Q = \frac{\text{max}(m_\pi, p)}{\Lambda_b}
\end{equation}
where $m_\pi$ is the pion mass, $p$ is a soft scale associated with the observable, and $\Lambda_b$ is a hard scale. In this work, we will set $\Lambda_b=600$ MeV and $p$ is given by the \NN{} scattering momentum. All contributions to the potential at chiral order $\nu=1$ vanish, i.e., $c_{1}=0$, since we employ Weinberg power counting. We refer to the different orders as \lo{} ($\nu=0$), \nlo{} ($\nu=2$), and \nnlo{} ($\nu=3$).

Truncating the \chieft{} expansion at order $k$ induces a truncation error given by
\begin{equation} \label{eq:truncation}
\delta y_\text{th}^{(k)} = y_\text{ref} \sum_{\nu = k+1}^K c_{\nu} Q^{\nu}.
\end{equation}
where $K \rightarrow \infty$. Assuming that \emph{all} $c_{\nu}$ coefficients, including those for which $\nu \leq k$, are independent and identically distributed, we can use known lower-order coefficients ${c_0,\ldots,c_k}$ to learn about the single \pdf\ from which the unknown higher-order coefficients should also be sampled. This, combined with a prior assumption about the form of the \pdf{} for the $c_{\nu}$ coefficients, provides us with a prescription for quantitatively estimating $\delta y_\text{th}^{(k)}$.

We assume a Gaussian prior with variance $\cbar^2$ for $c_\nu$, 
\begin{equation}
\prob(c_{\nu}|I) = \mathcal{N}\left(0, \bar{c}^2\right).
\end{equation}
All physical scales reside in $Q$ and $y_\text{ref}$ and it is reasonable to assume that the $c_{\nu}$ coefficients are of order $1$. This is an assumption we will test explicitly in Sec.~\ref{sec:estimating_cbar} when estimating $\bar{c}$ from order-by-order shifts in the prediction of different \NN{} scattering observables $y$.

Placing a normal prior for $c_{\nu}$, with known $\bar{c}^2$, leads to a normal \pdf{} with variance $\sigma_\text{th}$ for $\delta y_\text{th}^{\,(k)}$ in the limit $K\rightarrow \infty$ given by~\cite{wesolowski19}
\begin{equation} \label{eq:delta_y}
\prob(\delta y_\text{th}^{\,(k)} | \bar{c}^2, Q, I) = \mathcal{N}\left(0,
\sigma_\text{th}^2\right),
\end{equation}
where we have also conditioned on $Q$, and
\begin{equation}
\sigma_\text{th}^2 = \bar{c}^2 y_\text{ref}^2 \frac{Q^{2(k+1)}}{1 - Q^2}.
\end{equation}
In this work we will assume that the theory errors for different observables $(y_i, y_j)$ are completely uncorrelated. This leads to a diagonal covariance matrix $\Sigma_\text{th}$ for the truncation error at order $k$ that is given by
\begin{equation}
\label{eq:uncorr}
\left(\Sigma_\text{th}\right)_{ij} = \sigma_{\text{th},i}^2 \delta_{ij}.
\end{equation}
Note that the \lo{} truncation error is proportional to $Q^4 / (1-Q^2)$ since the $\nu=1$ chiral order vanishes such that the first non-zero term in Eq.~\eqref{eq:truncation} corresponds to $\nu=2$.

The likelihood function for the entire data set $\Dtrain$ can---assuming that experimental and theoretical errors are independent---be expressed as
\begin{equation}
\label{eq:likelihood}
\prob(\Dtrain|\lecs,I) \propto \exp(-\frac{1}{2} \vec{r}^{\,T} \cdot \left(\Sigma_\text{exp} + \Sigma_\text{th}\right)^{-1} \cdot \vec{r}),
\end{equation}
with the residual vector $\vec{r}$ defined as
\begin{equation}
\vec{r} = \vec{y}_\text{th}^{\,(k)}(\lecs) - \vec{y}_\text{exp},
\end{equation}
i.e., the difference between the experimental data and the corresponding set of \chieft{} predictions at order $k$ given a vector $\lecs$ of LEC values. We employ a diagonal covariance matrix $\Sigma_\text{exp}$ also for the experimental data and used the normalization factors from the Granada database~\cite{perez13-1, perez13-2} for the joint systematic uncertainty of a group of data originating from the same experiment~\cite{Stump:2001gu}. Note also that the likelihood~\eqref{eq:likelihood} is implicitly conditional on $\bar{c}$ and the breakdown scale $\Lambda_b$.

Let us also mention two possible extensions of the statistical model used in this work. Firstly, instead of placing an implicit delta prior on $\bar{c}^2$, one could follow standard Bayesian practice and exploit the class of conjugate priors, assuming, e.g., an inverse-$\chi^2$ \pdf{} such that $\prob(\bar{c}^2|I) = \chi^{-2}(\nu_0,\tau_0^2)$\footnote{Here $\nu_0$ and $\tau_0$ are hyperparameters corresponding to the degrees of freedom and scale for the $\chi^{-2}$ \pdf. This \pdf{} is equivalent to a normal-inverse-gamma \pdf.}. The conjugacy of this prior with respect to the normal \pdf{} leads to a Student's $t$ \pdf{} for the expansion parameters $c_\nu$ and the truncation error $\delta y_\text{th}^{(k)}$, see Ref.~\cite{Melendez:2019izc}. Secondly, one could relax the assumption of completely uncorrelated truncation errors and model the covariance structure of the expansion coefficients $c_{\nu}$ using, e.g., a Gaussian process where the correlation length in $Q$ is one of its hyperparameters~\cite{Melendez:2019izc}.


\subsubsection{Choice of reference values \label{sec:yref}}
The reference value, $y_\text{ref}$, in Eq.~\eqref{eq:truncation} for each observable can be assigned in different ways. Options include using an experimental value, picking some suitable theoretical prediction, or a value motivated from an order of magnitude estimate. We have chosen to let \lo{} predictions set the scale for observables. But since \lo{} predictions depend on the LECs $\lecs$---which at this point in the analysis remain to be inferred---it is necessary to extract a reasonable point estimate of \lo\ LEC values $\lecs_{\nu=0}^{\star}$. We do this by maximizing the likelihood \eqref{eq:likelihood} at \lo\ (i.e., setting $k=0$) using $\Sigma_\text{th}$ from Eq.~\eqref{eq:uncorr} with $y_\text{ref} = y_\text{exp}$ and $\bar{c} = 1$, as motivated by naturalness. The choice of using experimental data as reference values when maximizing the likelihood was technically convenient but has little practical impact on the subsequent steps of the analysis. In this way, we let the maximum likelihood estimate (MLE) at \lo, $y_\text{th}^{(0)}(\lecs_{\nu=0}^{\star})$, be the reference value for SGT and DSG observables. For spin polarization and correlation cross sections we use $y_\text{ref} = 0.15$ as motivated by the average of the \lo{} maximum-likelihood prediction.


\subsubsection{Estimating $\bar{c}$ \label{sec:estimating_cbar}}
We proceed to estimate $\bar{c}$ from the finite set of expansion coefficients $c_{\nu}$ obtained by rearranging Eq.~\eqref{eq:expansion} and setting $\lecs=\lecs_{\nu}^{\star}$. In general, we obtain a vector, $\vec{c}_{\nu}$, of expansion coefficients at order $\nu$
\begin{equation}
{c}_{\nu,i} = \frac{{y}_{\text{th},i}^{\,(\nu)}(\lecs_{\nu}^{\star}) - {y}_{\text{th},i}^{\,(\nu-1)}(\lecs_{\nu-1}^{\star})}{{y}_{\text{ref},i}  {Q}_i^{\nu}}, \quad i=1,\ldots,N_o,
\label{eq:cnu}
\end{equation}
for a representative set of $N_o=54$ observables at different scattering energies and angles. Specifically, we used a combination of total cross sections, differential cross sections, and spin observables in \np{} and \pp{} at an energy grid $T_\text{lab} = 20, 70, 120, 170, 220, 270$ MeV, and scattering angles $\theta = 50, 150$ degrees, such that we cover the relevant kinematical regions while being sufficiently well separated in the energy-angle variables to reduce the influence of finite correlations in the expansion coefficients. We account for the vanishing of chiral order $\nu=1$ when computing the order-by-order differences.

\input{tables/cbar}

At each order we then compute a root-mean-square (RMS) value, i.e.,
\begin{equation}
\bar{c}_{\nu} = \sqrt{\frac{1}{N_o} \sum_{i=1}^{N_o} c_{\nu,i}^2}.
\label{eq:crms}
\end{equation}
Our choice to set $y_\text{ref}$ according to the \lo{} MLE effectively constrains the expansion coefficient $c_0$, leading to $c_0=1.17$, accounting for the averaged spin polarizations. See the third column in Table~\ref{tab:cbar} for results at the other two orders. We also define an estimate of $\bar{c}$ as the total RMS value of the $c_{\nu,i}$ coefficients up to, and including, the order $k$ at which we truncate the EFT expansion (see the fourth column in Table~\ref{tab:cbar}). The highest-order estimate for $\bar{c}$ is given by $\bar{c}_{0 \ldots 3}$ and therefore includes information from \lo, \nlo, and \nnlo{}. Note that outlier values of $c_{\nu,i}$ were removed before computing all RMS values since they would otherwise influence our estimate disproportionally. We used the following procedure to determine outliers:
\begin{enumerate}
\item Compute the lower and upper quartiles $C_{25\%}$ and $C_{75\%}$ of $c_{\nu, i}$.
\item Determine the distance $\Delta C$ between the quartiles.
\item Discard any $c_{\nu,i}$ that falls outside the interval $\left[C_{25\%}-3 \Delta C, C_{75\%}+3 \Delta C\right]$.
\end{enumerate}
The numbers of removed outliers at each order are provided in Table~\ref{tab:cbar}.

Since we express the observable predictions as an EFT expansion in Eq.~\eqref{eq:expansion} we expect to observe expansion coefficients of natural size. On average we find relatively natural values that characterize the truncation error. However, the larger differences in the \chieft{} predictions when going from \lo{} to \nlo{} is a signature of an irregular convergence pattern and incurs an unexpectedly large value for $\bar{c}_{\nu=2}$. We analyze the consequences of this in Sec.~\ref{sec:model_checking}.


\section{Hamiltonian Monte Carlo}
Hamiltonian Monte Carlo shares some key features with the canonical Metropolis-Hastings algorithm~\cite{metropolis53, Hastings:1970aa}. These MCMC methods draw samples from a \pdf{} $\prob(\lecs)$ by producing an ergodic Markov chain of states whose unique stationary distribution is $\prob(\lecs)$. Given a current state $\lecscurr$ of the MCMC chain, the Metropolis-Hastings algorithm proposes a new state $\lecsprop$ from a proposal distribution $q(\lecsprop | \lecscurr)$. The new state is accepted with a probability $a$ given by
\begin{equation}
\label{eq:mh_acceptance_prob}
a = \text{min}(1, r)
\end{equation}
where $r$ is the Hastings ratio
\begin{equation}
\label{eq:hastings_ratio}
r = \frac{\prob(\lecsprop) q(\lecsprop | \lecscurr)}{\prob(\lecscurr)q(\lecscurr | \lecsprop)}
\end{equation}
The next state of the chain, $\lecs_{n+1}$, will be set to $\lecsprop$ if the update is accepted, or to a copy of $\lecscurr$ if it is rejected. The proposal distribution is typically a normal distribution but many other options exist. The Metropolis-Hastings algorithm is a single-particle (or single-walker) algorithm, i.e., it is only aware of a single location in the parameter space at any given time. The primary drawback of the random walk Metropolis-Hastings algorithm, and most of its derivatives, is that the elements of the resulting MCMC chain are strongly correlated with each other in most practical applications. Such correlations decrease the amount of information contained in the MCMC chain, requiring longer chains than would otherwise be necessary. The problem is exacerbated when the dimensionality of the parameter space for the sampled distribution $\prob(\lecs)$ increases.

In 1987, lattice field theorists devised an ingenious MCMC algorithm based on Hamiltonian dynamics to combat the problem of correlated samples. They dubbed the algorithm Hybrid Monte Carlo (HMC)~\cite{duane87}, but it has subsequently become known as Hamiltonian Monte Carlo~\cite{neal11}, retaining the acronym. HMC is based on the Metropolis-Hastings algorithm, but the method for proposing a new state is radically modified. The \pdf{} to be sampled is treated as a potential energy surface and the current state of the chain is regarded as a particle with position coordinates given by the parameters of the \pdf. By endowing the HMC ``particle'' at iteration $n$ with a randomly drawn momentum $\momcurr$ we can associate the joint state $(\lecscurr, \momcurr)$ with a total energy and, by extension, a probability to find the system in that state. Simulating the particle's trajectory for a finite period of time subsequently yields a proposed state $(\lecsprop, \momprop)$. The proposed state retains the previous total energy since this is a conserved quantity in Hamiltonian dynamics and the new state will therefore be accepted. The momentum $\momprop$ is then discarded. In practice, the total energy will only be approximately conserved due to numerical simulation errors, and we vet the proposed state in a manner similar to Eq.~\eqref{eq:mh_acceptance_prob}.

The result is a completely uncorrelated sample of the target \pdf, assuming that the length of the trajectory is appropriately chosen. Clearly, the length of the MCMC chain produced by HMC can then be drastically reduced compared to chains with a high degree of correlation. We will quantify this statement by extracting an effective number of samples in Sec.~\ref{sec:ess}. The strongest advantage of using the HMC algorithm is its potential for producing uncorrelated samples even when the target \pdf{} is high-dimensional. Unfortunately, simulating Hamiltonian dynamics for each proposed state is computationally expensive compared to the method of proposing new states by applying small random perturbations. It is therefore apparent that the reduction in overall chain length and inter-sample correlations must be sufficiently large to warrant the increased cost per sample.

In the following subsections we present the HMC algorithm in detail. We have opted to write a custom implementation\footnote{Available as free software at \url{https://github.com/svisak/montepython.git}.} using Python~\cite{van_rossum09} and NumPy~\cite{harris20} in lieu of using a standard package such as Stan~\cite{stan_manual21}. Specific implementation choices will be presented as appropriate. Our implementation and the mathematical details presented here are largely based on Ref.~\cite{neal11}. A conceptual introduction to HMC can be found in Ref.~\cite{betancourt18}.

\subsection{From potential energy to posterior probability}
In HMC we generate samples from the $d$-dimensional posterior \pdf{} $\prob(\lecs | \Dtrain, I)$ by simulating classical Hamiltonian dynamics for a particle moving under the influence of a potential proportional to the posterior itself. To see this, we let the phase space of states---specified by position and momentum coordinates $(\lecs,\mom\,)$ with total energy $H$---be described by a Boltzmann distribution
\begin{equation}
\prob(\lecs,\mom) = \frac{1}{Z}\exp(-H(\lecs,\mom)/T).
\label{eq:boltzmann}
\end{equation}
For a classical Hamiltonian function
\begin{equation}
H(\lecs,\mom) = K(\mom) + U(\lecs)
\label{eq:hmc_hamiltonian}
\end{equation}
with kinetic energy $K$ and potential energy $U$, the Boltzmann distribution factorizes and the marginal Boltzmann distribution for the position vector $\lecs$ is independent of the distribution for $\mom$. For our purposes, the temperature $T$ is some function that makes the exponent dimensionless, and we set $T=1$. The factor $Z$ is independent of $(\lecs,\mom\,)$ and ensures proper normalization of the distribution in Eq.~\eqref{eq:boltzmann}. From the exponential Boltzmann factor we identify the potential energy term as the negative log posterior (NLP). Disregarding the constant marginal likelihood we define
\begin{align}
  \begin{split}
\label{eq:potential_energy}
U(\lecs) &= - \log [\prob(\lecs|\Dtrain,I)] \\ &= -\log[\prob(\Dtrain|\lecs,I)\prob(\lecs|I)].
\end{split}
\end{align}
In the last step we also used Bayes' theorem \eqref{eq:bayes} to link the potential directly to the likelihood and the prior.

Following standard practice, and the classical dynamics analogy, we employ a quadratic form for $K(\mom)$:
\begin{equation} \label{eq:kinetic_energy}
K(\mom) = \frac{1}{2} \mom^{\, T} \MM^{-1} \mom.
\end{equation}
Here $\MM$ is a positive-definite, symmetric matrix, called the mass matrix. With $K(\mom) = - \log[\prob(\mom)]$ we have
\begin{equation}\label{eq:prob_momentum}
\prob(\mom\,) = \mathcal{N}(0,\MM).
\end{equation}
The HMC algorithm thus samples the joint \pdf{} $\prob(\lecs, \mom\,)$. Marginalizing over the auxiliary momentum $\mom$---by simply discarding these coordinates in the Markov chain---leaves us with MCMC samples of $\prob(\lecs|\Dtrain,I)$.

A particle governed by Hamiltonian dynamics moves through $2d$-dimensional phase space on a hypersurface of constant energy. The time evolution of the system is described by Hamilton's equations
\begin{align}
\frac{d\alpha_i}{dt} &= \frac{\partial H}{\partial p_i} \\
\frac{dp_i}{dt} &= -\frac{\partial H}{\partial \alpha_i}
\end{align}
where $i = 1 \ldots d$. The form of the Hamiltonian, Eq.~\eqref{eq:hmc_hamiltonian}, allows us to rewrite Hamilton's equations as
\begin{align}
\label{eq:hamilton1}
\frac{d\alpha_i}{dt} &= \frac{\partial K}{\partial p_i} = (\MM^{-1}p)_i \\
\label{eq:hamilton2}
\frac{dp_i}{dt} &= -\frac{\partial U}{\partial \alpha_i}.
\end{align}
Note that $\partial U / \partial \alpha_i$ is a partial derivative of the NLP that we must evaluate in order to simulate Hamiltonian dynamics and, by extension, use HMC sampling. Realistically, this must be done through automatic differentiation (AD) except in special cases where analytic expressions for these partial derivatives are available (our implementation readily allows for both possibilities). AD generally incurs a factor of two overhead compared to just evaluating the target \pdf~\cite{Griewank:2003}. In this work, we exploit an external AD library~\cite{charpentier09} for computing the necessary gradients and we measure the computational overhead of AD to less than 50\%, see Sec.~\ref{sec:correlated_samples}.


\subsection{Advancing the HMC sampler} \label{sec:hmc_sampler}
Starting from some current state $(\lecscurr,\momcurr)$, the total energy of a particle trajectory traversing the HMC phase space is conserved. We should thus always accept the proposed state $(\lecsprop,\momprop)$ at the end of the Hamiltonian trajectory, discard the auxiliary momentum $\momprop$, and store $\lecsprop$ as the new parameter sample $\lecs_{n+1}$ in the HMC chain. In practice, however, numerical errors in the solution of Eqns.~\eqref{eq:hamilton1}-\eqref{eq:hamilton2} break energy conservation, which in turn breaks detailed balance. If ignored, we can no longer guarantee that the Markov chain converges to the sought stationary distribution. The solution is to vet the proposed state, i.e., the state at the end of the Hamiltonian trajectory, with the accept/reject step of the Metropolis-Hastings algorithm. We can also ensure a symmetric proposal distribution by negating the momentum variable $\momprop$ at the end of the particle trajectory. However, such a negation is not necessary for a quadratic momentum distribution as in Eq.~\eqref{eq:prob_momentum}. The new state is thus accepted with the probability $a$ given by Eq.~\eqref{eq:mh_acceptance_prob} and a Hastings ratio for the joint probabilities given by
\begin{equation}
r = \frac{\prob(\lecsprop, \momprop)}{\prob(\lecscurr, \momcurr)} = \exp\left(-H(\lecsprop, \momprop) + H(\lecscurr, \momcurr)\right).
\end{equation}
The HMC chain will typically be ergodic since we draw a new momentum before integrating Hamilton's equations and thereby drastically altering the total energy.


\subsection{Leapfrogging Hamiltonian dynamics}
\label{sec:leapfrog}
Upon imposing the Metropolis-Hasting accept/reject criterion we implicitly require reversibility of our chain. Fortunately, Hamiltonian dynamics is time-reversible and it is necessary to integrate Hamilton's equations in a way that preserves this property. Standard methods like Euler integration or explicit Runge-Kutta methods are disqualified as they do not preserve time-reversibility. From a strictly practical point of view, integration methods that do not conserve the Hamiltonian also limit the length of the particle trajectory since a large accumulated error would result in an unacceptably low mean acceptance rate $\accrate$. To ensure time-reversibility, one should use a symplectic integrator that goes hand-in-hand with the volume preservation of phase space that follows from Liouville's theorem. The local discretization error of a symplectic integrator is equally likely to be positive or negative in each step of the integration as long as the step size $\epsilon$ is below some threshold value. The result is that the total energy is approximately conserved for an arbitrarily long trajectory. A nice property of HMC is that $\accrate$ drops precipitously if $\epsilon$ is greater than the threshold value. The absence of quiet failures makes it trivial to diagnose a too large choice for $\epsilon$. The upper bound for $\epsilon$ is generally imposed by the most constrained parameter in the posterior.

The number of leapfrog iterations $L$ can drastically influence the performance of the HMC sampler; it is imperative that $L$ is set neither too high nor too low. A too small number partly defeats the purpose of using HMC in the first place, as it would result in a random-walk-like behavior with highly correlated samples. In contrast, a too large value of $L$ would waste valuable CPU cycles without improving (and possibly even curtailing) performance. Naturally, choosing a very small step size $\epsilon$ needs to be compensated for by increasing $L$ in order to avoid random walks.

Neither $\epsilon$ nor $L$ are fixed in our implementation. Rather, random values are drawn from predefined probability distributions prior to each invocation of the leapfrog solver according to
\begin{align}
\epsilon &\sim \mathcal{U}\left(\frac{1}{2} \epsilon^{\star}, \frac{3}{2} \epsilon^{\star}\right) \\
L &\sim \mathcal{U}\left\{\frac{1}{2} L^{\star}, \frac{3}{2} L^{\star}\right\}, \quad \frac{1}{2} L^{\star} \geq 1,
\end{align}
where the user specifies the nominal values $\epsilon^{\star}$ and $L^{\star}$. The reasons for randomizing these leapfrog parameters are threefold. First, variations in the trajectory length $\epsilon L$ may decrease correlations between samples. Second, a fixed trajectory length can result in oscillatory behavior if $\epsilon L$ happens to approximately match some periodicity of the target distribution. This type of (nearly) non-ergodic behavior can severly limit the efficiency of HMC. Third, the target \pdf{} may have regions where its gradient is very steep so that the nominal $\epsilon$ is too large to resolve features in that section.

There are extensions of HMC whose purpose is to relieve the user from the burden of tuning the hyperparameters $\epsilon$ and $L$. For instance, the choice of $\epsilon$ may be automated based on acceptance rates of small trial runs. The state-of-the-art No-U-Turn Sampler (NUTS) terminates trajectories based on heuristic rules for when continued simulation no longer increases the performance of the sampler. Both of these improvements are described in Ref.~\cite{homan14}.


\subsection{Tuning in to the target distribution}
An array of tunable HMC parameters have been introduced in the previous subsections: the step size $\epsilon$, the number of leapfrog iterations $L$, and the mass matrix $\MM$. A drawback of HMC is the need to carefully tune these hyperparameters to each target distribution, or risk poor performance. However, as mentioned in Sec.~\ref{sec:leapfrog}, the tuning of $\epsilon$ and $L$ may be automated. For now, we use a manual tuning procedure that will be outlined below. It is designed to achieve efficient sampling and also involves the important mass matrix.

The keen reader may have noticed that $\epsilon$, $L$, and $\MM$ are interlinked and changing one may force us to change one (or both) of the others. The tuning procedure is therefore, to an extent, iterative. We start with $\epsilon$.


\subsubsection{Leapfrog step size and number of iterations} \label{sec:tuning_epsilon_L}
For tuning $\epsilon$ we exploit that the acceptance rate $\accrate$ is largely independent of $L$ (a consequence of using a symplectic integrator) and produce a very short HMC chain using a small number of leapfrog steps ($L = 3$). If $\accrate$ is low (in practice 0\%) we decrease $\epsilon$ by an order of magnitude and try again. If $\accrate$ is 100\% we instead increase $\epsilon$ by an order of magnitude and try again. Once the appropriate magnitude is found we make fine-grained adjustments as necessary. Using this method we quickly achieved HMC  acceptance rates of 99\% at all three chiral orders. This value is likely too high for optimal efficiency, but we find it adequate for our purposes.

The number of required leapfrog iterations $L$ is intimately linked to the choice of the mass matrix $\MM$. Careful tuning of $L$ is rather pointless until $\MM$ is settled. We have found that $L \approx$ 10--20 yields excellent performance for approximately Gaussian distributions with around a dozen or so parameters, assuming that the mass matrix (and $\epsilon$) is well chosen. To rapidly assess the choice of $L$---and the overall performance---it is useful to inspect trace plots of each individual parameter. The trace plots will reveal no apparent structures if the HMC algorithm is performing well. Any remaining structures in the trace plots may be quenched by increasing $L$, at the obvious expense of increased computational effort, or by improving the mass matrix. The latter alternative should always take precedence if possible. Note that it is usually necessary to revisit the tuning of $\epsilon$ after the mass matrix has been updated.


\subsubsection{Mass matrix}
Both $\epsilon$ and $L$ are scalar values with no distinction for each individual parameter $\alpha_i$ and thus cannot be used to compensate for differences in parameter scales. Like Metropolis-Hastings, and unlike, e.g., affine-invariant ensemble samplers (AIES)~\cite{goodman10} such as the \emcee\ package~\cite{foremanmackey13}, HMC is sensitive to such differences of scale and we need a way to account for them. This is the purpose of the mass matrix $\MM$.

Deploying a mass matrix that captures the most important features of the target distribution is absolutely critical to the performance of HMC. An improper choice of $\MM$ can degrade the performance by several orders of magnitude. Letting $\MM = \mathbb{1}$, i.e., an identity matrix, does not fare well with the \chieft{} models analyzed in this work. To improve, we exploit published LEC uncertainties from a previous analysis~\cite{reinert18} and construct a diagonal mass matrix. We then draw $\sim 1000$ samples, using $L = 8$ at \lo{} and $L = 20$ at \nlo\ and \nnlo, to estimate a parameter covariance matrix $\Sigma_{\lecs}$ and construct a mass matrix according to $\MM = \Sigma^{-1}_{\lecs}$. We find that this approach to learn about $\MM$ yields high HMC performance in practice. Note that one does not have to use HMC for this tuning, as we do; indeed, it may be preferable to use a more expedient method for extracting an approximate parameter covariance matrix, e.g. the more tuning-agnostic MCMC sampler \emcee{}.

Local estimates of the target covariance based on, e.g., optimization and second derivatives have been used in previous studies of LEC uncertainties, see, e.g., Ref.~\cite{Carlsson:2015vda}. This method was recently used to construct a Bayesian prior for the \chieft{} contact LECs at \nnlo{} when estimating the $c_D$ and $c_E$ LECs in the three-nucleon force sector~\cite{wesolowski21}. It was found that the prior and marginal posterior were largely the same. This finding reinforces the observation that the inverse of a point-estimated covariance matrix yields a performant mass matrix.


\section{Sampling LEC posteriors using Hamiltonian Monte Carlo}
\label{sec:lec_posteriors}
In Sec.~\ref{sec:HMC_strategy} we outline the sampling strategy and in Sec.~\ref{sec:LEC_posteriors} we present our posterior \pdf s $\prob(\lecs|\Dtrain,I)$ for the LECs $\lecs$ at \lo, \nlo, and \nnlo\ in \chieft\ sampled using HMC. These posteriors enable all subsequent inference in this paper and constitute the main result of our work. In Sec.~\ref{sec:convergence} we discuss the convergence of the MCMC chains. In Sec.~\ref{sec:ess} we highlight some of the unique aspects of the HMC algorithm by comparing with posterior samples obtained using \emcee{}. In Sec.~\ref{sec:multimodality} we comment on multimodality and the challenge it brings.

\input{tables/chain_stats}


\subsection{Sampling strategy \label{sec:HMC_strategy}}
We employ the same HMC sampling strategy at all chiral orders considered in this work.

\begin{enumerate}
\item To identify a ballpark region where we expect to find the posterior mode we first optimize the data likelihood in Eq.~\eqref{eq:likelihood}. At this stage we employ $\bar{c}=1$ and $y_\text{ref}=y_\text{exp}$ to parameterize the covariance matrix for the EFT truncation error. Every subsequent HMC sampling is then randomly initiated within an overdispersed region around the MLE.
\item To tune the mass matrix $\MM$, we use previously published uncertainties for the LECs~\cite{reinert18} to define its diagonal entries and draw $n_\text{tune} \sim 1000$ samples using $L \approx$ 10-20. The resulting sample covariance matrix is inverted to yield the final mass matrix.
\item We determine the optimal step size $\epsilon$ from a small set of very short HMC chains consisting of 10--20 samples and easily find a step size that yields an acceptance rate $\accrate$ of 99\%.
\item Equipped with a well-tuned HMC algorithm, we collect $M \geq 3$ independent chains at different starting values for the LECs. It is important to have two or more chains to enable canonical convergence tests based on within- and between-chain variances, e.g., the Gelman-Rubin test. We also employ a convergence criterion based on the integrated autocorrelation time as recommended in Ref.~\cite{foremanmackey13,Sokal}, see Sec.~\ref{sec:correlated_samples}.
\item The trace plots of the HMC chains indicate that the length of the burn-in phase is very short. This is corroborated by the autocorrelation analysis presented in Sec.~\ref{sec:correlated_samples}. We discard the initial $\sim 10$ samples from each chain, except for three chains at \nlo\ which require us to discard $\sim 100$ samples. The situation is drastically different for the \emcee{} chains, where we find it necessary to discard the initial $\sim 1000$ samples at \lo\ and $\sim$ 30,000--40,000 samples at the higher orders. These lengthy burn-ins have to be repeated for each \emcee\ chain. In contrast, tuning the HMC hyperparameters is a one-time cost.
\end{enumerate}

Detailed information about the tuning and sampling phases at \lo--\nnlo{} are summarized in Table~\ref{tab:chain_stats}. Note that the number of tuning samples $n_\text{tune}$ is larger than necessary at \lo\ and \nlo. The step sizes $\epsilon$ and acceptance rates $\accrate$, which are closely linked, are remarkably similar across the three chiral orders. The reason is that the step size is generally limited by the most constrained parameter(s) which in all three cases are the \lo\ LECs $\Ct$.


\subsection{LEC posteriors \label{sec:LEC_posteriors}}
The posterior \pdf s for the LECs are multivariate but will be presented using univariate and bivariate projections. These so-called corner plots of the LO, NLO, and NNLO posteriors are shown in Figs. \ref{fig:lo_glob_corner}, \ref{fig:nlo_glob_corner}, and \ref{fig:n2lo_glob_corner}, respectively.

At all orders, the locations of the maximum a posteriori (MAP) probability and widths of the posterior \pdf s are similar to the corresponding measures we obtained using frequentist parameter estimation in Ref.~\cite{Carlsson:2015vda}. This is largely due to the fact that we are using nearly the same database of thousands of \NN{} scattering cross sections in both analyses and that the inference is likelihood dominated. The main modification to the database comes from setting aside part of the data for validation in this work. We also employ identically regulated $\chi$EFT interactions, and closely related diagonal covariance matrices for estimating uncorrelated EFT truncation errors. Despite several apparent similarities it is very important to realize that we are comparing results from two fundamentally different approaches. The use of Bayesian inference methods allows us to assign a probability (density) measure to LEC values themselves. In the frequentist approach we are estimating covariances from the gradients at the maximum likelihood estimator of the \emph{data}.


\subsubsection{\lo}
At LO we consider the two \NN{} contact LECs present at this order: $\Ct_{1S0}$ and $\Ct_{3S1}$, acting in the $S$-waves. The corner plot in Fig.~\ref{fig:lo_glob_corner} reveals that they are both very well constrained by the \NN{} scattering data $\Dtrain$ and appear to be uncorrelated with each other. We note that $\Ct_{1S0}$ is considerably more constrained than $\Ct_{3S1}$. This is likely due to: (i) the isovector (isoscalar) character of $\Ct_{1S0}$ ($\Ct_{3S1}$), and( ii) that \pp\ data is more abundant and more precise than \np\ data at low scattering energies where the truncation error is relatively small.
\input{texfig/lo_glob_corner}
%

\subsubsection{\nlo}
\input{texfig/nlo_glob_corner}

Several contact LECs are introduced at \nlo{} and we find that the LEC posterior exhibits noticeable correlations in certain directions, see Fig.~\ref{fig:nlo_glob_corner}. The presence of such correlations indicate a level of parameter redundancy in the model. From a statistical perspective there exists methods, e.g., singular value decomposition, to identify and retain only the most important parameters (or linear combinations of parameters) of a model to explain data, so-called stiff directions in the parameter space. However, before doing so it is worthwhile to inspect the model structure from a physics perspective. In the present case we identify a strong correlation between the LECs $\Ct_{1S0}^{pp}$ and $\Ct_{1S0}^{np}$. Following conventional counting of the isospin-breaking effects in \chieft\ we encounter the leading isospin-dependent $^1S_0$ contacts at \nlo{}. This is also in line with the results from a high-precision data analysis by the Nijmegen group~\cite{Bergervoet:1988zz} demonstrating that strong and electromagnetic interactions break charge independence and most prominently in the $^{1}S_0$ channel. However, since isospin-breaking is a comparatively small effect, a non-negligible EFT truncation error at \nlo{} is likely to dilute isospin sensitivity with respect to the \NN{} data being used.

We also detect correlations between $S$-wave LECs acting within the same spin channel. In general, the spin-singlet and spin-triplet partial-wave contact LECs do not exhibit any significant correlation with each other at any of the chiral orders we examine. This is somewhat different from the frequentist analysis in Ref.~\cite{Carlsson:2015vda} where correlations ($|\rho|\gtrsim 0.7$) where found between all $S$-wave LECs and $C_{3P2}$. We speculate that this difference in correlation structure could be rooted in the difference between the models of the truncation error used in the frequentist and Bayesian analyses. Upon inspection, we find that the truncation errors at \nlo{} and \nnlo{} employed in this work are more than twice as large compared to the corresponding error magnitudes used in Ref.~\cite{Carlsson:2015vda}. We have not performed a systematic analysis to compare the two error models, but a smaller truncation error can certainly shift weights of, e.g., spin-polarization and spin-averaged data at different energies which in turn could induce stronger correlations between the LECs of the interaction model.

\input{texfig/n2lo_glob_corner}
\subsubsection{\nnlo}
At \nnlo{} we also find strong correlations between certain LECs, see Fig.~\ref{fig:n2lo_glob_corner}. Similarly to the \nlo{} posterior, there is a very strong correlation between the charge-independence breaking contacts. However, we do not identify any unexpected correlation structures, like the ones found at \nnnlo{} in Ref.~\cite{wesolowski19}, to reveal a physics parameter redundancy in the model.

At this order we encounter the sub-leading \piN{} LECs $c_1,c_3,c_4$ for the first time. As opposed to contact LECs they act in all angular momentum channels. Certain combination of \piN{} and \NN{} LECs also show significant correlations; in particular, $c_1$ appears positively correlated with the $\Ct_{1S0}$ LECs, as does $c_3$ with, e.g., $C_{3P2}$.

Recall that we assign a multivariate normal prior \pdf{} for $c_1,c_3,c_4$ using maximum-likelihood results from a Roy-Steiner analysis of the $\Delta$-less \piN{} scattering amplitudes~\cite{siemens17}. This prior strongly regulates the values of $c_1,c_3,c_4$ compared to the \NN{} LECs which are assigned a less informative prior based on naturalness.

In Fig.~\ref{fig:n2lo_glob_piN_contours} we show a corner plot where we focus on the differences between the prior and posterior in the \piN-sector. Note that we are comparing two \pdf s that have very different origin from both a statistical and methodological perspective. We can still draw the following conclusions:
\begin{itemize}
\item The \NN{} data induces an overall $5-10$\% shift, in the positive direction, of the \piN{} LECs, i.e., the \NN{} data appears to reduce the \piN{} sub-leading attraction slightly. However the overall effect of this shift on the binding of atomic nuclei remains to be analyzed.
\item The difference in MAP values of the \piN{} prior and posterior is significant compared to the extent of the credible intervals. For $c_1$ we observe a slight overlap of the marginal \pdf s. It is interesting to note that the \piN{} vertex corresponding to $c_1$ does not contain any contribution from the $\Delta$-isobar.
\item The link between low-energy \piN{} and \NN{} scattering processes is a hallmark of $\chi$EFT. If \emph{all} uncertainties are accurately modeled, and we are operating with an EFT, then we expect to find an overlap between the prior and the posterior. However, we do not observe this. One possible explanation is that we estimate the truncation error in the \NN\ sector via the uncorrelated theory covariance matrix in Eq.~\eqref{eq:uncorr}. The Roy-Steiner analysis is based on an MLE with uncorrelated data and method uncertainties~\cite{Hoferichter:2015hva,siemens17} and it is not clear how to propagate an EFT error to the \piN\ LECs matched at this order.
\end{itemize}
\input{texfig/n2lo_glob_piN_contours}
%

\subsection{Convergence towards a stationary distribution \label{sec:convergence}}
How many samples do we need to reach an accurate representation of the stationary target distribution with small sampling error? In connection with this, one should note that the $N$ samples collected during some finite time period will not be independent.

It is unfortunately not possible to determine the level of convergence of a finite chain; we can only attempt to detect convergence failures. As such, all convergence diagnostics in the MCMC literature merely provide necessary but not sufficient conditions. Multiple diagnostics have been devised, of which we employ two of the most common ones: the standard Gelman-Rubin statistic ($\hat{R}$)~\cite{gelman92,brooks98}, discussed in Sec.~\ref{sec:gelmanrubin}, and the integrated autocorrelation time ($\tau$) discussed in Sec.~\ref{sec:correlated_samples}. Both diagnostics are applied to each LEC $\alpha_i$ individually.

Although MCMC algorithms are ergodic and eventually explores the entire state space, we clearly face the challenge of pseudo-convergence due to multimodality when working with finite chains. This problem is discussed further in Sec.~\ref{sec:multimodality}.


\subsubsection{The Gelman-Rubin statistic $\hat{R}$ \label{sec:gelmanrubin}}
With the Gelman-Rubin statistic $\hat{R}$ we compare the variance of the samples within a single chain to the variance between $M \geq 3$ chains initialized at different starting positions. Following Ref.~\cite{gelman92}, we assume that each chain contains $N$ samples after we have discarded initial samples to reduce the memory of the starting position. We discussed the removal of the burn-in samples in Sec.~\ref{sec:HMC_strategy}. Based on the $N \times M$ samples one defines a joint mean
\begin{equation}
\bar{\alpha}_i = \frac{1}{M}\sum_{m=1}^M \bar{\alpha_i}^{(m)}
\end{equation}
for each LEC $\alpha_i$ based on all chains where $\bar{\alpha_i}^{(m)}$ denotes the within-chain mean for the $m$th chain and is given by
\begin{equation}
\label{eq:within_chain_mean}
\bar{\alpha}_i^{(m)} = \frac{1}{N}\sum_{n=1}^N \alpha_i^{(nm)}.
\end{equation}
In this notation, $\alpha_{i}^{(nm)}$ corresponds to the $n$th sample of the $i$th LEC in the $m$th MCMC chain. One can express the between-chain and within-chain variances for the $i$th LEC in terms of the corresponding means as
\begin{equation}
B_i = \frac{N}{M-1}\sum_{m=1}^{M} \left(\bar{\alpha}_i^{(m)} - \bar{\alpha}_i \right)^2
\end{equation}
and
\begin{equation}
 W = \frac{1}{M}\sum_{m=1}^M \frac{1}{N-1}\sum_{n=1}^N \left( \alpha_i^{(nm)} - \alpha_i^{(m)}\right)^2,
\end{equation}
respectively.

\input{texfig/gelman_rubin}

A weighted average of the above variances can be used to estimate the variance of the marginal posterior for the LEC $\alpha_i$
\begin{equation}
\textnormal{Var}^{+}[\alpha_i] = \frac{N-1}{N} W + \frac{1}{N}B
\end{equation}
where the $+$-sign indicates that this quantity overestimates the posterior variance provided that the $M$ chains are initialized at locations with greater variability compared to the true posterior. Indeed, for finite $N$ we have that $W$ will underestimate the marginal variance since the chains have not explored the posterior while $B$ will overestimate the marginal variance if the initial sampling distribution is overdispersed. In the limit of $N \rightarrow \infty$ we will have that $W$ approaches the variance of the marginal posterior. Incorporating these finite-$N$ corrections leads to a Student's $t$ distribution for $\alpha_i$ with variance (scale) estimated by
\begin{equation}
 V = \textnormal{Var}^{+}[\alpha_i] + \frac{B}{MN}.
\end{equation}
The Gelman-Rubin measure expresses the potential scale reduction by forming the ratio
\begin{equation}
\hat{R} = \sqrt{\frac{V}{W}},
\end{equation}
which approaches 1 as $N \rightarrow \infty$. A widely used threshold for declaring convergence is $\hat{R}<1.01$~\cite{bda3}. This, somewhat arbitrary, threshold simply states that the sample variance is 2\% larger than the within-chain variance, and that one should expect a corresponding potential scale-reduction if continuing the sampling process. In Fig.~\ref{fig:gelman_rubin} we show the evolution of $\hat{R}$ with the number of HMC samples at \lo{}, \nlo{}, and \nnlo{}. Clearly, our HMC chains fulfill $\hat{R}<1.01$ as well as an even stricter threshold $\hat{R} < 1.001$. The chains obtained with \emcee{} also pass the same $\hat{R}$ thresholds after a similar amount of MCMC samples.

There exists updated Gelman-Rubin measures. One can employ so-called split-$\hat{R}$~\cite{bda3} and rank-normalized $\hat{R}$~\cite{Vehtari2021} to better handle non-stationary chains and chains distributed with a heavier tail that conspire to yield a good $\hat{R}$. We have not detected the need for using such updated $\hat{R}$ measures to analyze the \chieft{} posteriors in this work.


\subsubsection{The integrated autocorrelation time $\tau$ \label{sec:correlated_samples}}
We use the MCMC chains to compute (statistical) expectation values and we should therefore also analyze the sampling variance of such estimates. For example, one can straightforwardly estimate the mean value $\bar{\alpha}_i$ of the $i$th LEC within a single MCMC chain\footnote{In this section we discuss quantities pertaining to a single chain and therefore omit the superscript $(m)$.}, see Eq.~\eqref{eq:within_chain_mean}. The sampling variance of the estimated mean value for a particular LEC, based on $N$ uncorrelated samples, scales with $1/N$ according to
\begin{equation}
\textnormal{Var}[\bar{\alpha}_i] \equiv \textnormal{E}[\bar{\alpha}_i - \textnormal{E}(\alpha_i)] = \frac{\textnormal{Var}[\alpha_i]}{N},
\label{eq:variance_mean}
\end{equation}
where $\textnormal{Var}[\alpha_i]$ is the variance of the samples with respect to the posterior $\prob(\alpha_i|\Dtrain,I)$. This estimate only holds for uncorrelated samples, and we will demonstrate that such samples can be obtained with the HMC algorithm. In contrast, most random-walk based MCMC algorithms generate highly correlated samples. When the samples are correlated, the variance of the mean is modified according to
\begin{equation}
\textnormal{Var}[\bar{\alpha}_i] = \tau_i \frac{\textnormal{Var}[\alpha_i]}{N}
\label{eq:MC_variance}
\end{equation}
where $\tau_i$ is referred to as the integrated autocorrelation time for the chain of sample values of the $i$th LEC. It is given by
\begin{equation}
\tau_i = \lim_{N \to \infty} \left(1 + 2\sum_{h=1}^{N} \rho_i(h)\right).
\label{eq:integrated_autocorrelation}
\end{equation}
The autocorrelation function $\rho_i(h)$ measures the correlation between (stationary) samples separated by $h$ MCMC steps.
\input{texfig/acors_comparison}
In the literature, $h$ is referred to as the \emph{lag}. Note that the integrated autocorrelation time will be different for each expectation value. Here, we limit ourselves to inspect $\tau_i$ for the mean values of the LECs and use this below to assess the convergence of an MCMC chain. In Fig.~\ref{fig:acors_comparison} we present the estimated autocorrelation functions $\hat{\rho}_i(h)$ of all LECs at LO, NLO, and NNLO, and their averages, as obtained by HMC and \emcee{}. As expected, a well-tuned HMC algorithm generates virtually uncorrelated samples whereas the \emcee{} chains exhibit a correlation structure that is typical for most MCMC algorithms. The HMC algorithm generates uncorrelated samples even as the dimensionality of the parameter space is increased. We see this advantageous dimensionality-scaling of HMC when going from \lo{} to \nnlo{}. The corresponding correlation length of the \emcee{} chains markedly increase. This is one of the primary advantages of using HMC.

Some care is needed in the numerical computation of the integrated autocorrelation time $\tau_i$. For large values of $N$ in Eq.~\eqref{eq:integrated_autocorrelation}, the estimated autocorrelation $\hat{\rho}$ suffers from a signal-to-noise problem. Indeed, although the correlation decreases towards zero with increasing lag, the variance of the correlation does not. Following Ref.~\cite{Sokal} we therefore truncate the sum at the smallest integer $N^{\star}$ such that $N^{\star} \geq c \tau_i(N^{\star})$ for $c=5$. With this in place we can monitor the evolution and convergence of $\tau$ as a function of the number of collected samples $N$ in the chain.

As mentioned, the computation of $\tau$ is not only useful for quantifying the sampling variance but also provides a handle on the convergence of the MCMC chain. While the Gelman-Rubin statistic compares several identically prepared chains, diagnosing convergence based on the evolution of $\tau$ can be applied to a single chain. Using this method, convergence is declared when the estimation of $\tau$ has stabilized and $N \gg \tau$, where $N$ is the length of the chain. In this work we apply the condition $N \geq 50\tau$, and in Fig.~\ref{fig:tau_vs_nsamples} we show the evolution of $\tau$ as $N$ increases. We have also indicated the non-convergence zone $N < 50\tau$. At \lo{} and \nlo{}, we fulfill $\tau$-convergence using both HMC and \emcee{}. At \nnlo{}, we fulfill $\tau$-convergence only with HMC while \emcee{} falls just short of the imposed tolerance.

\input{texfig/tau_vs_nsamples}


\subsection{Effective sample size and efficiency of HMC \label{sec:ess}}
The reduction in sample quality due to correlation is often quantified with the \emph{effective sample size}
\begin{equation}
\textnormal{ESS} = \frac{N}{\tau},
\label{eq:ess}
\end{equation}
where we also introduce $\tau$ defined as the average of all $\tau_i$ for each LEC $\alpha_i$. We use the ESS value to quantitatively compare the efficiencies of HMC and \emcee{}. It is obvious from Eqns.~\eqref{eq:MC_variance} and~\eqref{eq:ess} that the correlation structure directly impacts the ESS and hence the total computational effort required to reach a tolerable variance. Since there is a significant computational overhead involved in integrating Hamilton's equations, it is key to tune the HMC algorithm to reach a very small $\tau$. The intrinsic benefit of using HMC also increases with the dimensionality of the parameter space.
 
\input{tables/ess}

In Table~\ref{tab:ess} we present $\tau$ values and relevant related quantities for \emcee{} and HMC at \lo{}, \nlo{}, and \nnlo{}. The results presented in this table are based on a single chain for each order and choice of algorithm. We have, however, verified that the results are similar regardless which one of the parallel chains that is used. We note that we have $\tau \approx 1$ for the HMC sampled chains, and even $\tau < 1$ in two cases (\lo\ and \nlo). With \emcee, $\tau$ is roughly two orders of magnitude greater than the HMC equivalents. This is solely due to the correlation structure of the respective MCMC chains. Also shown in Table~\ref{tab:ess} (as ESS/$N$) is the average number of effective samples that one MCMC sample provides. By introducing $N_\mathcal{L}$---the total number of likelihood calls during sampling, tuning, and burn-in---we also show (as $\evalpersample$) the average number of calls to the likelihood function required to generate one MCMC sample. We obtain \evalpersample\ $> 1$ also for \emcee\ since the quoted results also include burn-in. We can use the number of likelihood calls per effective sample to compare the average efficiency of HMC and \emcee, as evaluating the likelihood constitutes nearly all of the computational effort. We therefore define an HMC speedup factor $\mathcal{S}$ with respect to \emcee{} as
\begin{equation}
\mathcal{S} = \frac{[\evalperess]_\text{\emcee}}{[\evalperess]_\text{HMC} \times [\textnormal{AD-cost}]},
\label{eq:speedup}
\end{equation}
where we also account for the computational overhead induced by the use of AD which we employ to generate the gradients necessary for HMC. In our implementation we measure the AD overheads to 10\% at \lo, 24\% at \nlo, and 43\% at \nnlo, and we use these figures when quantifying the HMC speedup factors in Table~\ref{tab:ess}. In summary we find that the real-world speedup factor is more than 6 at \lo\ and \nlo, and 3.6 at \nnlo. The smaller speedup at \nnlo\ is primarily due to less ideal tuning, and could be improved further. Another contributing factor is that the estimate of $\tau$ for \emcee\ at \nnlo\ has not stabilized, and the value reported in Table~\ref{tab:ess} is therefore a lower bound (cf. Fig.~\ref{fig:tau_vs_nsamples}).

We obtain $\tau<1$ at \lo{} and \nlo{} when using HMC, a result which comes from drawing anticorrelated samples from the posterior \pdf, as can be seen clearly in the corresponding autocorrelation functions in Fig.~\ref{fig:acors_comparison}. This so-called antithetic sampling~\cite{hammersley_morton_1956} leads to ESS $> N$, i.e., an effective number of samples greater than the number of MCMC samples with a corresponding reduction of the variance in Eq.~\eqref{eq:MC_variance}. Therefore, drawing completely independent MCMC samples is not necessarily the optimal strategy. In this work we encounter antithetic sampling for sufficiently large values of the HMC step length $L$ once we have constructed a mass matrix $\mathcal{M}$ that we believe suits the target distribution. This is obviously an advantageous sampling strategy and antithetic sampling is one of several known variance reduction techniques in Monte Carlo sampling~\cite{neal11}. While we do not achieve antithetic sampling at \nnlo\ in this work, we believe that it can be reached with further tuning.


\subsubsection{Consequences of improper tuning of the HMC hyperparameters}
HMC sampling is challenging in practice, primarily due to the need for careful tuning of the mass matrix $\MM$ to achieve high performance. It is therefore instructive to show what a failure looks like. Fig.~\ref{fig:nat_pub_acors} demonstrates the strong autocorrelation that results when $\MM$ is not well-tuned for two chains at \nlo; one where $\MM$ is set up using a \chieft{} naturalness argument, and the other where it is based on previously published LEC uncertainties~\cite{reinert18}. To construct the mass matrix using the naturalness argument we employed ratios of the contact LECs at \lo--\nnlo{} according to
\begin{equation}
\abs{\Ct_i} \sim \frac{4\pi}{F_\pi^2}, \quad \abs{C_i} \sim \frac{4\pi}{F_\pi^2\Lambda_b^2}
\end{equation}
where $F_\pi \simeq 92$ MeV is the pion decay constant. In both cases, the integrated autocorrelation time is very large which, combined with the high per-sample computational cost of HMC, results in very poor performance. Autocorrelation structures like those shown in Fig.~\ref{fig:nat_pub_acors} are unacceptable of course, but typically seen before tuning the hyperparameters.

\input{texfig/nat_pub_acors}


\subsection{Multimodality
\label{sec:multimodality}}
Although the Markov chain is ergodic, one would have to run the MCMC algorithm for an infinitely long time to visit all states. Clearly, multimodal distributions makes the sampling process considerably more complicated. The Markov chain has to cross a valley of low probability to explore more than one mode. Such a crossing is naturally a low probability event and will generally occur infrequently. Different modes may also have different shapes, causing poor performance if the tuning of the MCMC algorithm is unsuitable for the other mode even if the algorithm successfully moves between modes. Standard HMC has no advantage over alternative sampling algorithms in this regard due to its single-walker nature combined with the necessity of careful problem-specific tuning. Although the situation can be improved with, e.g., tempering methods~\cite{neal11} it is probably better to explore other MCMC algorithms, such as MultiNest~\cite{10.1111/j.1365-2966.2009.14548.x}, if multimodality is expected.

\input{texfig/lo_scan}

In our analysis we encountered one clear case of multimodality: the LEC posterior at \lo{}. This \pdf{} is straightforward to explore in detail because it is only two-dimensional. We performed a scan on a $500 \times 500$ grid of the \lo{} \pdf{}. The result is shown in Fig.~\ref{fig:lo_scan}. The main mode is located around $(\Ct_{1S0},\Ct_{3S1}) \approx (-0.11,-0.07) \cdot \Ctunit$, and marked with a cross in the lower panel of Fig.~\ref{fig:lo_scan}. A second mode is found at $(\Ct_{1S0},\Ct_{3S1}) \approx (-0.11,-0.03) \cdot \Ctunit$. We computed the marginal likelihood $\prob(D|I)$ of both modes using the Laplace approximation, and found that the second mode contains a negligible probability mass. A deep valley---many orders of magnitude lower in probability---separates the two modes and presents an effectively impenetrable barrier to the HMC sampler. The valley is caused by the breakup of the deuteron bound state, at $\Ct_{3S1} \approx -0.05 \cdot \Ctunit$, that suppresses the likelihood for \np\ SGT data at very low energies. As far as we can tell the vast majority of the probability mass for the LEC posteriors at \nlo{} and \nnlo{} is located in a single dominant mode. In the current work we therefore proceed under the assumption that all \pdf s are unimodal.


\section{Model checking}
\label{sec:model_checking}
A probabilistic model of a physical system can only be upheld if it provides an acceptable representation of data. Therefore, we should always check to what extent the model fits the data. To that end we sample the \ppd s and inspect the empirical coverages.

\subsection{The posterior predictive distribution}
As presented in Sec.~\ref{sec:data}, we have reserved roughly one third of the Granada database of experimentally measured scattering cross sections for model validation and refer to this as data set $\Dval$, see Table~\ref{tab:databases}.

Following Eq.~\eqref{eq:ytrue} we can model the true value of an \NN{} scattering observable as the independent sum of the predicted value up to chiral order $k$ and the truncation error, i.e.,
\begin{equation}
y_\text{true} = y_\text{th}^{(k)} + \delta y_\text{th}^{(k)}.
\label{eq:ypredict}
\end{equation}
Neither term on the right hand side is known with certainty. Indeed, they are stochastic variables described by \pdf s. We consider the prediction $y_\text{th}^{(k)}$ uncertain due to the uncertainty of the LECs, and we consider $\delta y_\text{th}^{(k)}$ uncertain due to the unknown EFT expansion coefficients $c_{\nu}$ for $\nu > k$. Additional uncertainties regarding, e.g., the breakdown scale $\Lambda_b$, the expansion parameter $Q$, and the correlation structure of the model predictions $y_\text{th}^{(k)}$, can be accounted for as well~\cite{wesolowski21}. In this work, however, we focus on quantifying the LEC uncertainty and combining this with the EFT truncation error.

Equipped with an HMC chain of samples from the LEC posterior $\prob(\lecs|\Dtrain,I)$ at chiral order $k$, it is straightforward to evaluate the ppd $\text{pr}(y_\text{th}^{(k)} | \Dtrain, I)$ by observing that
\begin{align}
  \begin{split}
\text{pr}(y_\text{th}^{(k)} | \Dtrain, I) &= \int \prob(y_\text{th}^{(k)}, \lecs | \Dtrain, I) \,
d\lecs \\ &= \int \text{pr}(y_\text{th}^{(k)} | \lecs, \Dtrain, I)
\text{pr}(\lecs | \Dtrain, I) \, d\lecs \\ &= \int
\text{pr}(y_\text{th}^{(k)} | \lecs, I) \text{pr}(\lecs | \Dtrain, I) \,
d\lecs.
  \end{split}
\end{align}
The last step is a consequence of the conditional independence between $y_\text{th}^{(k)}$ and $\Dtrain$ given $\lecs$. Drawing random samples from this ppd amounts to evaluating $y_\text{th}^{(k)}$ for each of the $N$ samples $(\vec{a}^{(1)},\vec{a}^{(2)}, \ldots, \vec{a}^{(N)})$ in the HMC chain. Fortunately, the HMC chains are comparatively short and the necessary computation of \NN{} scattering observables does not pose any significant challenge. Should this become an issue one could try emulating the observable response~\cite{Konig:2019adq,Ekstrom:2019lss,melendez2021fast} or use hardware acceleration~\cite{miller2021accelerating}. Opting for emulation or acceleration will add an error term to Eq.~\eqref{eq:ypredict} quantifying the corresponding additional uncertainty. %

\input{texfig/ppd_validation_385_215}

Next, we sample the \pdf{} for the truncation error in Eq.~\eqref{eq:delta_y}. This is trivial for a normally distributed EFT error where the parameters $\bar{c}^2$ and $y_\text{ref}$ characterize this \pdf{} entirely.

Since we can evaluate the two terms in Eq.~\eqref{eq:ypredict} we can also draw samples from $\prob(y_\text{true}|D,I)$. Panels (a) and (b) in Fig.~\ref{fig:ppd_validation_385_215} show such predictions for the true value of the total \np{} cross section (SGT) for $0 < \Tlab \leq 350$ MeV, while panels (c) and (d) show predictions of the true value for the \pp{} spin correlation parameter (AYY) at $\Tlab = 294.4$ MeV. These observables were not included in the training data set $\Dtrain$. Also shown in the figure are experimental measurements of the same observables, gathered from Refs.~\cite{lisowski82} for SGT and \cite{vonPrzewoski:1998ye} for AYY. The panels in the left column show 100 individual predictions at each order, while the panels in the right column show 68\% and 95\% highest density intervals (HDIs)\footnote{For a unimodal \pdf, a $p \%$ HDI will be the smallest interval around the mode of the \pdf{} that comprises $p \%$ of its probability mass. } computed from 1,000 predictions. We only find unimodal \ppd s and they appear to be rather symmetric. Indeed, as will be discussed below, the \ppd s are dominated by the normally distributed EFT truncation error. The individual predictions were generated in an uncorrelated fashion to reflect our model for the EFT truncation error. See the supplemental material~\cite{suppl} for \ppd s at \lo{}, \nlo{}, and \nnlo{} of all observables present in the Granada database, i.e., a model check with respect to both the training data $\Dtrain$ and the validation data $\Dval$.

The predictions in Fig.~\ref{fig:ppd_validation_385_215} appear to converge toward the experimental results with increasing chiral order, and a visual inspection of the 68\% and 95\% HDIs of the \ppd s indicates that they work as advertised for these observables. We note that the employed model for the truncation error does not incorporate the known symmetry constraints of the spin-scattering matrix~\cite{Hoshizaki:1969qt}, e.g., that the vector analyzing power P goes to zero at extreme scattering angles. This type of information can be straightforwardly incorporated as a boundary condition on a Gaussian process model for the EFT truncation error~\cite{Melendez:2017phj}.

We find that the truncation error $\delta y_\text{th}^{(k)}$ is the dominating source of uncertainty in Eq.~\eqref{eq:ypredict}. Indeed, the propagated error due to the variability in the LECs is quite small in comparison since the LEC posteriors are conditioned on a very large and informative data set. In fact, we find that the truncation error dominates at all orders up to \nnlo{} for all energies in the Granada database. Even at low energies ($\Tlab \approx 1$ MeV) the truncation error is still 5--10 times larger than the uncertainty stemming from the LECs. Of course, at low energies both errors are very small on an absolute scale. In accordance with EFT principles we would expect the HDIs to narrow with each order and widen with increasing scattering energy. Instead, we find that the widths of the HDIs at all orders are comparable, and particularly so at the larger $T_\text{lab}$-values. This is a consequence of underestimating $\bar{c}$ at \lo{} and \nlo{}, see Table~\ref{tab:cbar}. We remind the reader that we estimate the EFT truncation at order $k$ by exploiting information only up to this order since this reflects what one can do in a real situation without any higher orders available. Of course, at \lo{} the available information is particularly scarce. In Sec.~\ref{sec:covprob} we will explore other choices for estimating $\bar{c}$.


\subsection{Empirical coverage probability} \label{sec:covprob}
\input{texfig/empirical_coverage}
We compute the empirical coverage probability to assess whether our HDIs are accurate as advertised. The empirical coverage quantifies how well we meet the expectation that if we compute a $p \cdot 100$~\% HDI for the ppd of a true value for an observable, we should find that this HDI covers the measurement of said observable with probability $p$ on average. To that end, we perform a binary test for each datum in our validation data set $\Dval$ and count the number of times the validation data falls within the specified HDI, i.e., we count the number of `hits', and compare this with the total number of data in $\Dval$. In this procedure we neglect the (generally) very small uncertainties in the experimental data. Repeating the binary coverage test for a range of values of $p$, i.e., for different HDI intervals, yields a summary for the empirical coverage probability that is also convenient to inspect graphically.

Figure \ref{fig:empirical_coverage} shows the resulting empirical coverage plots for the validation data set $\Dval$ using three different strategies for evaluating an RMS value for $\cbar$. In coverage plots of this kind, a $p \cdot 100$~\% HDI should yield a coverage such that it ends up on the 45\textdegree\ positive diagonal if it is working as expected. The diagonal line is indicated with a dash-dotted line in all our coverage plots. A coverage probability larger than this nominal value, i.e., a coverage probability above the diagonal, corresponds to an overly wide HDI, implying a too conservative error. The opposite situation corresponds to an overly narrow (liberal) HDI, implying a too small assigned error. Following a precautionary principle, a conservative error is preferable to the underestimated one.

The observed number of hits should also follow a binomial distribution under the assumption that the observables are uncorrelated, see, e.g., Ref.~\cite{furnstahl15}. Our validation set, like the training set, contains hundreds of largely independent data groups that contain data recorded at different experimental facilities over several decades. The continuous version of the binomial distribution is the $\beta$ distribution, so we use this to compute 95\% confidence intervals, indicated as a gray filled region along the diagonal in all coverage plots.

The truncation error is intimately linked to the inferred value for $\cbar$. In Fig.~\ref{fig:empirical_coverage}(a) we estimate $\bar{c}$ by the RMS-value $\bar{c}_k$ according to Table~\ref{tab:cbar}. This method entails that we use information from all orders up to the one we are working at, i.e., $\bar{c}_k = \text{RMS}(\vec{c}_{0 \ldots k})$. Clearly, this yields liberal HDIs for the \ppd s on average, in particular at \lo\ and \nnlo. Note that the \lo{} HDI is based on an EFT truncation error essentially governed by the choice of reference values. At \nlo{} and \nnlo{}, the HDIs represent credible intervals conditioned on information from one and two order-by-order differences, respectively, and the corresponding coverage probabilities perform slightly better. However, we clearly underestimate $\cbar$.

As a second approach to determine $\cbar$, which we denote $\text{RMS}(\vec{c}_{2 \ldots k})$, we ignore the $\bar{c}_{\nu=0}$ contribution. In this case the empirical coverage probability improves drastically, see Fig.~\ref{fig:empirical_coverage}(b). Motivated by the dominance of the EFT truncation error compared to the uncertainty originating from LEC variability, we do not resample the LEC posteriors when varying the hyperparameters of the truncation error. The \lo{} predictions are absent in Fig.~\ref{fig:empirical_coverage}(b) to emphasize that this method for computing $\cbar_k$ is not relevant at that order.

In Fig.~\ref{fig:empirical_coverage}(c) we compute $\cbar$ as the RMS value of the expansion coefficients extracted with respect to the first omitted order only, which we denote $\text{RMS}(\vec{c}_{k+1})$. At \nnlo{} we exploit an MLE at \nnnlo{}. As expected, this method improves the performance of our error model, in particular at \lo{} and \nlo{}. There is not much difference between the coverage probabilities for the \nnlo{} HDI in panels (b) and (c).

\input{texfig/empirical_coverage_subsets}

In Fig.~\ref{fig:empirical_coverage_subsets} we inspect the coverage probability with respect to three different subsets of the validation data set $\Dval$. In panel (a), we only look at integrated \np{} cross section data (\np\ SGT). In (b), we look at all data with $\Tlab \leq 100$ MeV, and in (c) we look at all data with $\Tlab > 290$ MeV. Here we use $\bar{c}_k = \text{RMS}(\vec{c}_{2 \ldots k})$. The coverage with respect to SGT data is consistently too high. The data points in this set are very correlated with each other, which makes this comparison less meaningful. In panel (b), where we retain validation data with $\Tlab \leq 100$ MeV, the error model at \nlo{} performs rather well, whereas the \nnlo{} error is too liberal. The coverage with respect to data with $T_\text{lab}>290$ MeV, panel (c), exhibits a similar pattern. Although the HDIs improve with information from higher orders, as expected, this is not an entirely satisfactory situation when one is making predictions at low orders of \chieft{}.

Finally, we estimate the consequences of modifying the employed value for the \chieft{} breakdown scale $\Lambda_b$. Specifically, we change $\Lambda_b$ from 600 MeV to 500 MeV in the expression for the EFT truncation error \eqref{eq:delta_y} without resampling the LEC posteriors. We find that decreasing $\Lambda_b$ results in more natural $\cbar$ values while simultaneously yielding improved coverage probabilities for the EFT error. The model also becomes somewhat less sensitive to the method of computing $\cbar$. Conversely, raising $\Lambda_b$ increases the model's sensitivity to $\cbar$ as well as resulting in less naturally sized expansion coefficients. The results presented here are in line with the findings in Ref.~\cite{Melendez:2017phj}. Further studies of the \chieft\ breakdown scale, along the lines of Ref.~\cite{wesolowski21}, are clearly warranted.


\section{Conclusions and outlook}
In this work we implemented the HMC MCMC algorithm for sampling the LEC posterior \pdf{} $\prob(\lecs|\Dtrain,I)$ at \lo, \nlo, and \nnlo{} in the \NN{} sector of \chieft{}. We accounted for uncorrelated EFT truncation errors~\cite{furnstahl15} in the sampling. Our prior was based on both reasonable assumptions and information from previous studies. For example, we assume natural EFT expansion coefficients and \NN{} contact LECs while the $\pi N$ sector was further informed by the results from a Roy-Steiner analysis~\cite{siemens17} of $\pi N$ scattering data.

We conditioned the LEC posteriors on thousands of scattering data, leading to a likelihood-dominated posterior and consequently a probability mass of the LEC posterior that is concentrated to a very small region in parameter space. At all orders, the MAP and typical widths of the posteriors are very close to the corresponding measures found using frequentist parameter estimation in, e.g., Refs.~\cite{Carlsson:2015vda, Reinert:2017usi}. This is largely due to the fact that we employed nearly the same database of \NN{} scattering cross sections in both analyses.

An analysis of the coverage probability of the HDIs for predictions indicates that the credible intervals perform largely as advertised if one excludes the \lo\ expansion parameter when estimating $\bar{c}$ and if one has access to predictions at \nnlo\ or beyond. Indeed, the large shift in predictions when going from \lo\ to \nlo\ does not provide representative information about the EFT convergence pattern. This disturbance in the order-by-order description of the nuclear interaction was also identified in a recent Bayesian analysis of three- and four-nucleon states~\cite{wesolowski21}. It is hence desirable and timely to develop improved models for the EFT truncation error that explicitly account for such irregularities in the convergence pattern. At the same time it is equally important to address deficiencies in the LO description of the nucleon-nucleon interaction, in particular for making reliable low-order EFT predictions~\cite{Yang:2020pgi}.

Apart from a second mode in the \lo{} posterior, for which the deuteron is unbound, we found no clear evidence of multimodality in the LEC posteriors at \nlo{} and \nnlo. We also found that the posterior for the $\pi N$ LECs at \nnlo{} does not overlap with any significance with the narrow prior inferred from the Roy-Steiner analysis in Ref.~\cite{siemens17}. Provided that we are operating with a low-energy EFT for the nuclear interaction, it is reasonable to expect this discrepancy to vanish if all uncertainties are accurately modeled.

We found that the HMC algorithm provides virtually uncorrelated samples of the LEC posterior at all three chiral orders. We confirm a very low level of autocorrelation which is a hallmark of the HMC algorithm. When analyzing the correlation structure in detail we found evidence of an antithetic sampling pattern in the HMC chains at chiral orders \lo\ and \nlo. This yields an integrated autocorrelation time $\tau <1$ and an ESS greater than the number of gathered MCMC samples. If harnessed, antithetic sampling could serve as a valuable method for reducing the sample variance of LEC posteriors at higher chiral orders.

We also compared the HMC chains at the three considered orders to corresponding chains obtained using the \emcee\ algorithm and found a three- to six-fold increase of the sampling efficiency. With HMC, we achieved near-instant convergence, as measured by $\tau$, at \lo, \nlo, and \nnlo{}. Using \emcee, in contrast, we just managed to cross the convergence threshold at \nlo{} and failed to reach a converged \nnlo{} result. Although HMC sampling relies on access to gradients of the posterior with respect to the LECs, and requires more tuning than other MCMC algorithms, we find that the rewards justify the extra effort. It is certainly possible to devise algorithms for self-tuning of the hyperparameters with NUTS-HMC~\cite{homan14} being one such strategy. Our results provide a promising outlook for parameter estimation at higher chiral orders, where the increasing number of LECs most likely presents a formidable challenge to other MCMC algorithms.


\begin{acknowledgments}
We thank Martin Hoferichter and Jacobo Ruiz de Elvira for supplying the central values and full covariance matrices for the $\pi$N LECs. This work was supported by the European Research Council (ERC) under the European Unions Horizon 2020 research and innovation programme (Grant agreement No. 758027), the Swedish Research Council (Grant No. 2017-04234). The computations were enabled by resources provided by the Swedish National Infrastructure for Computing (SNIC) at Chalmers Centre for Computational Science and Engineering (C3SE), the National Supercomputer Centre (NSC) partially funded by the Swedish Research Council.
\end{acknowledgments}



\bibliography{manuscript}

\end{document}

%% file: in/macros.tex
\newcommand{\cbar}{\bar{c}}
\newcommand{\lecs}{\vec{\alpha}}
\newcommand{\chieft}{$\chi$EFT}
\newcommand{\lo}{LO}
\newcommand{\nlo}{NLO}
\newcommand{\nnlo}{NNLO}
\newcommand{\nnnlo}{N3LO}
\newcommand{\NN}{$NN$}
\newcommand{\piN}{$\pi N$}
\newcommand{\pp}{$pp$}
\newcommand{\np}{$np$}

\newcommand{\Ctunit}{
  \ifmmode 10^4\text{ GeV}^{-2} \else $10^4$ GeV$^{-2}$ \fi
}
\newcommand{\Cunit}{
  \ifmmode 10^4\text{ GeV}^{-4} \else $10^4$ GeV$^{-4}$ \fi
}
\newcommand{\cunit}{
  \ifmmode ~\text{GeV}^{-1} \else GeV$^{-1}$ \fi
}

\newcommand{\Ct}{\widetilde{C}}
\newcommand{\MM}{\mathcal{M}}

\newcommand{\accrate}{\bar{a}}

\newcommand{\evalpersample}{
 \ifmmode N_{\mathcal{L}}/N \else $N_{\mathcal{L}}/N$ \fi
}
\newcommand{\evalperess}{
 \ifmmode N_{\mathcal{L}}/\textnormal{ESS} \else $N_{\mathcal{L}}/$ESS \fi
}

\newcommand{\pdf}{pdf}
\newcommand{\ppd}{ppd}
\newcommand{\emcee}{\texttt{emcee}}
\newcommand{\prob}{\text{pr}}

\newcommand{\lecscurr}{\lecs_n}
\newcommand{\lecsprop}{\lecs_n'}
\newcommand{\momcurr}{\mom_n}
\newcommand{\momprop}{\mom\,'_n}

\newcommand{\mom}{\vec{p}}
\newcommand{\Dval}{\widetilde{D}}
\newcommand{\Dtrain}{D}

\newcommand{\Tlab}{T_{\text{lab}}}


%% file: tables/databases.tex
\begin{table}[htpb]
\caption{The distribution of observables in the training and validation data sets denoted $\Dtrain$ and $\Dval$, respectively. $\Dval$ is composed of all available data in the $80 \leq T_\text{lab} \leq 100$ and $290 < T_\text{lab} \leq 350$ MeV ranges, plus one set of $np$ SGT data~\cite{lisowski82} covering the $33 \leq T_\text{lab} \leq 350$ MeV range. All other available data constitutes $\Dtrain$. We denote scattering observables using the SAID nomenclature~\cite{SAID}, e.g., integrated cross section (SGT), unpolarized differential cross section (DSG), polarization (P), polarization of the beam (PB), and polarization of the target (PT). These observables make up the bulk part of the experimental \NN{} database.}
\begin{ruledtabular}
\begin{tabular}{ c c c c c c c }
 & \multicolumn{3}{ c }{Training data ($\Dtrain$)} & \multicolumn{3}{ c }{Validation data ($\Dval$)} \\
\colrule
Obs & \np & \pp & Total (\%) & \np & \pp & Total (\%) \\
\colrule
SGT & 315 & 0 & 315 (7.2) & 84 & 0 & 84 (4.2) \\
SGTL & 11 & 0 & 11 (0.3) & 4 & 0 & 4 (0.2) \\
SGTT & 16 & 0 & 16 (0.4) & 3 & 0 & 3 (0.1) \\
DSG & 1221 & 756 & 1977 (45.2) & 457 & 159 & 616 (30.5) \\
A & 5 & 47 & 52 (1.2) & 0 & 24 & 24 (1.2) \\
AP & 0 & 5 & 5 (0.1) & 0 & 0 & 0 (0.0) \\
AT & 30 & 0 & 30 (0.7) & 35 & 0 & 35 (1.7) \\
AXX & 0 & 143 & 143 (3.2) & 0 & 120 & 120 (5.9) \\
AYY & 64 & 151 & 215 (4.8) & 46 & 151 & 197 (9.8) \\
AZX & 0 & 137 & 137 (3.1) & 0 & 120 & 120 (5.9) \\
AZZ & 45 & 39 & 84 (1.9) & 27 & 10 & 37 (1.8) \\
CKP & 0 & 1 & 1 (0.0) & 0 & 1 & 1 (0.0) \\
D & 13 & 56 & 69 (1.6) & 14 & 42 & 56 (2.8) \\
D0SK & 8 & 0 & 8 (0.2) & 14 & 0 & 14 (0.7) \\
DT & 39 & 0 & 39 (0.9) & 39 & 0 & 39 (1.9) \\
MSKN & 0 & 8 & 8 (0.2) & 0 & 8 & 8 (0.4) \\
MSSN & 0 & 8 & 8 (0.2) & 0 & 8 & 8 (0.4) \\
NNKK & 8 & 0 & 8 (0.2) & 0 & 0 & 0 (0.0) \\
NSKN & 12 & 0 & 12 (0.3) & 13 & 0 & 13 (0.6) \\
NSSN & 4 & 0 & 4 (0.1) & 14 & 0 & 14 (0.7) \\
P & 0 & 489 & 489 (11.2) & 0 & 260 & 260 (12.9) \\
PB & 590 & 0 & 590 (13.5) & 253 & 0 & 253 (12.5) \\
PT & 38 & 0 & 38 (0.9) & 19 & 0 & 19 (0.9) \\
R & 5 & 50 & 55 (1.3) & 0 & 50 & 50 (2.5) \\
RP & 0 & 22 & 22 (0.5) & 0 & 5 & 5 (0.2) \\
RPT & 1 & 0 & 1 (0.0) & 1 & 0 & 1 (0.0) \\
RT & 29 & 0 & 29 (0.7) & 37 & 0 & 37 (1.8) \\
\colrule
All & 2454 & 1912 & 4366 (100) & 1060 & 958 & 2018 (100) \\
\end{tabular}
\end{ruledtabular}
\label{tab:\currfilebase}
\end{table}

%% file: tables/cbar.tex
\begin{table}
\caption{Results from the $\bar{c}$ analysis. $\bar{c}_{\nu}$ denotes the RMS value of the EFT expansion coefficients at a particular order $\nu$, while $\bar{c}_{0 \ldots \nu}$ denotes the RMS value up to, and including, order $k$.}
\begin{ruledtabular}
\begin{tabular}{ c c c c c c }
Order & $\nu$ & $\bar{c}_{\nu}$ & $\bar{c}_{0\ldots\nu} $ & Outliers (order $\nu$) & Outliers (all orders) \\
\colrule
\lo{}   & 0 & 1.17 & 1.17 & 1/54 & 1/54 \\
\nlo{}  & 2 & 4.95 & 2.08 & 0/54 & 15/108 \\
\nnlo{} & 3 & 2.84 & 2.72 & 1/54 & 8/162 \\
\end{tabular}
\end{ruledtabular}
\label{tab:\currfilebase}
\end{table}

%% file: tables/chain_stats.tex
\begin{table*}
\caption{Detailed statistics of the HMC chains during the tuning and sampling phases. The number of chains at each order is $M$ and the total number of samples is counted across all $M$ chains. The HMC parameters $\epsilon^{\star}$ and $L^{\star}$ denote the nominal step length and total number of steps taken with the leapfrog algorithm to integrate Hamiltons equations for each HMC step. $\accrate$ is the average acceptance rate.}
\begin{ruledtabular}
\begin{tabular}{ c c c c c c c }
Order & $n_\text{tune}$ & $M$ & Total number of samples & $\epsilon^{\star}$ & $L^{\star}$ & $\accrate$ \\
\colrule
\lo & 2,000 & 3 & 50,063 & 0.1 & 8 & 99\% \\
\nlo & 2,000 & 10 & 57,134 & 0.09 & 20 & 99\% \\
\nnlo & 591 & 3 & 10,155 & 0.08 & 20 & 99\% \\
\end{tabular}
\end{ruledtabular}
\label{tab:\currfilebase}
\end{table*}

%% file: texfig/lo_glob_corner.tex
\begin{figure}[b]
\centering
\includegraphics[width=1.0\linewidth]{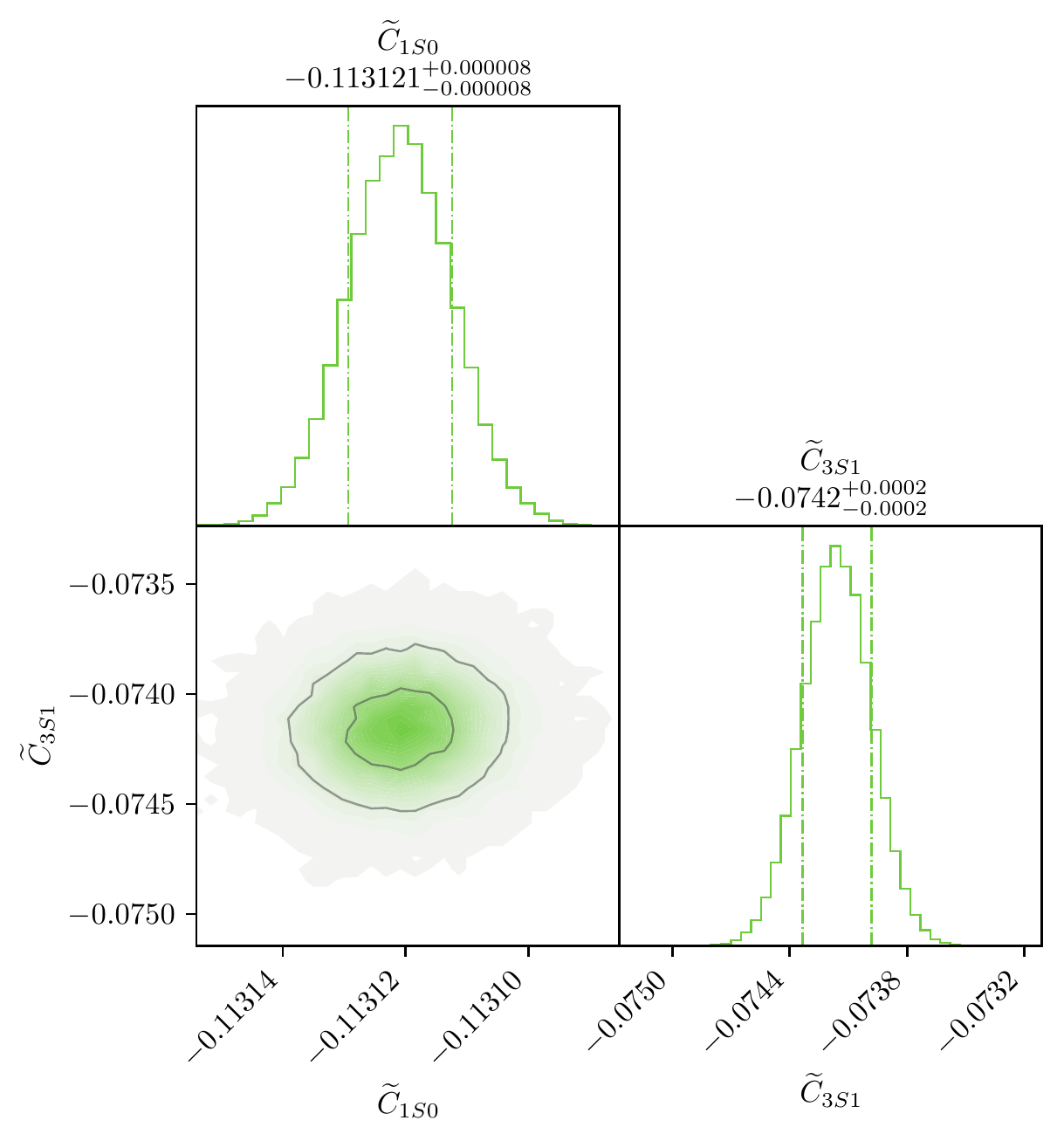}
\caption{\lo\ posterior sampled with HMC. The LECs are shown in units of $\Ctunit$. The inner (outer) gray contour line encloses 39\% (86\%) of the probability mass. The dot-dashed vertical lines indicate a 68\% credibility interval in the univariate marginals. White areas indicate zero counts of samples.}
\label{fig:\currfilebase}
\end{figure}

%% file: texfig/nlo_glob_corner.tex
\begin{figure*}[htpb]
\centering
\includegraphics[width=1.0\linewidth]{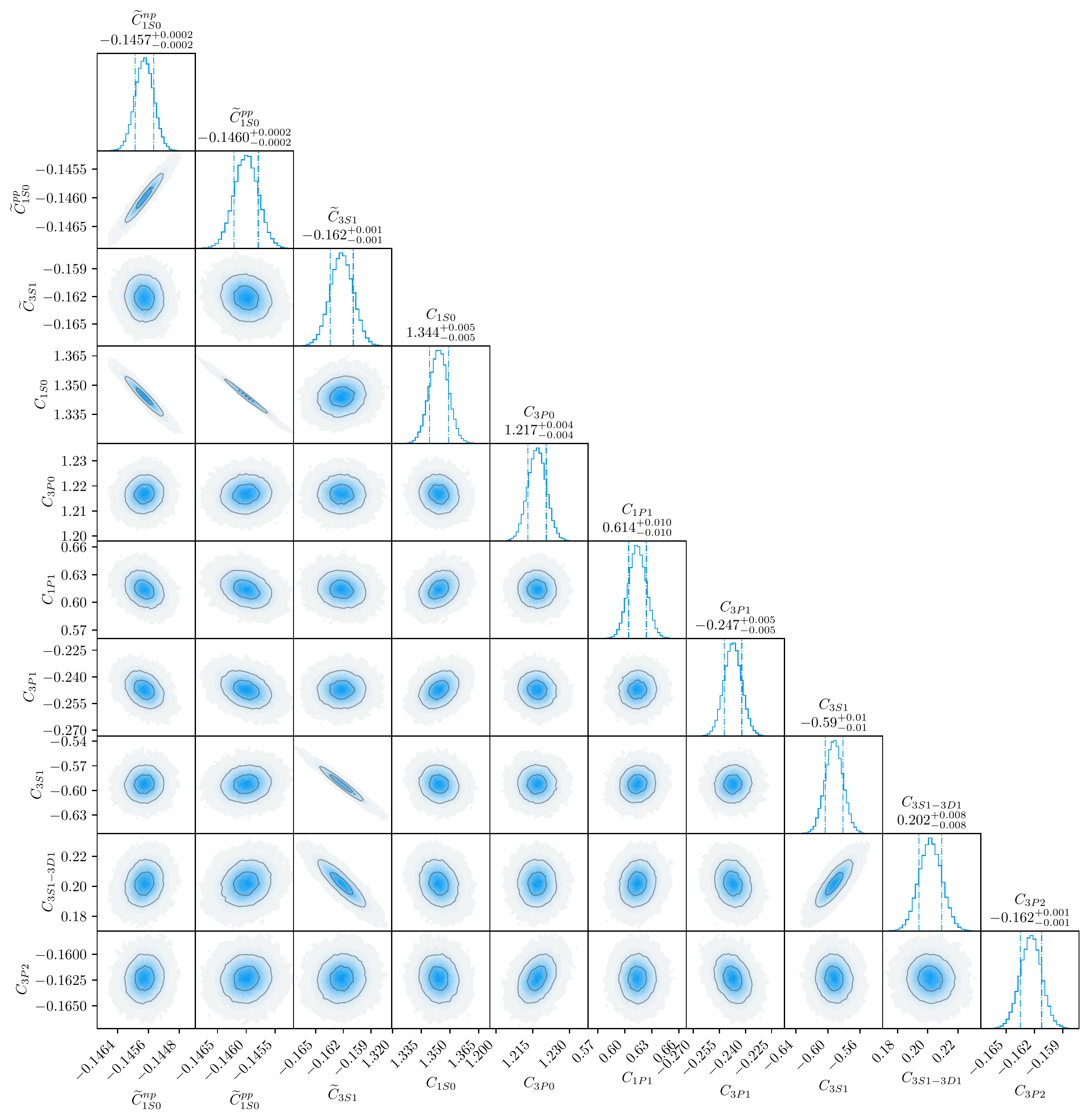}
\caption{\nlo\ posterior sampled with HMC. The LECs are shown in units of $\Ctunit$ for the \lo\ LECs and $\Cunit$ for the \nlo\ LECs. The inner (outer) gray contour line encloses 39\% (86\%) of the probability mass. The dot-dashed vertical lines indicate a 68\% credibility interval in the univariate marginals.}
\label{fig:\currfilebase}
\end{figure*}

%% file: texfig/n2lo_glob_corner.tex
\begin{figure*}[htpb]
\centering
\includegraphics[width=1.0\linewidth]{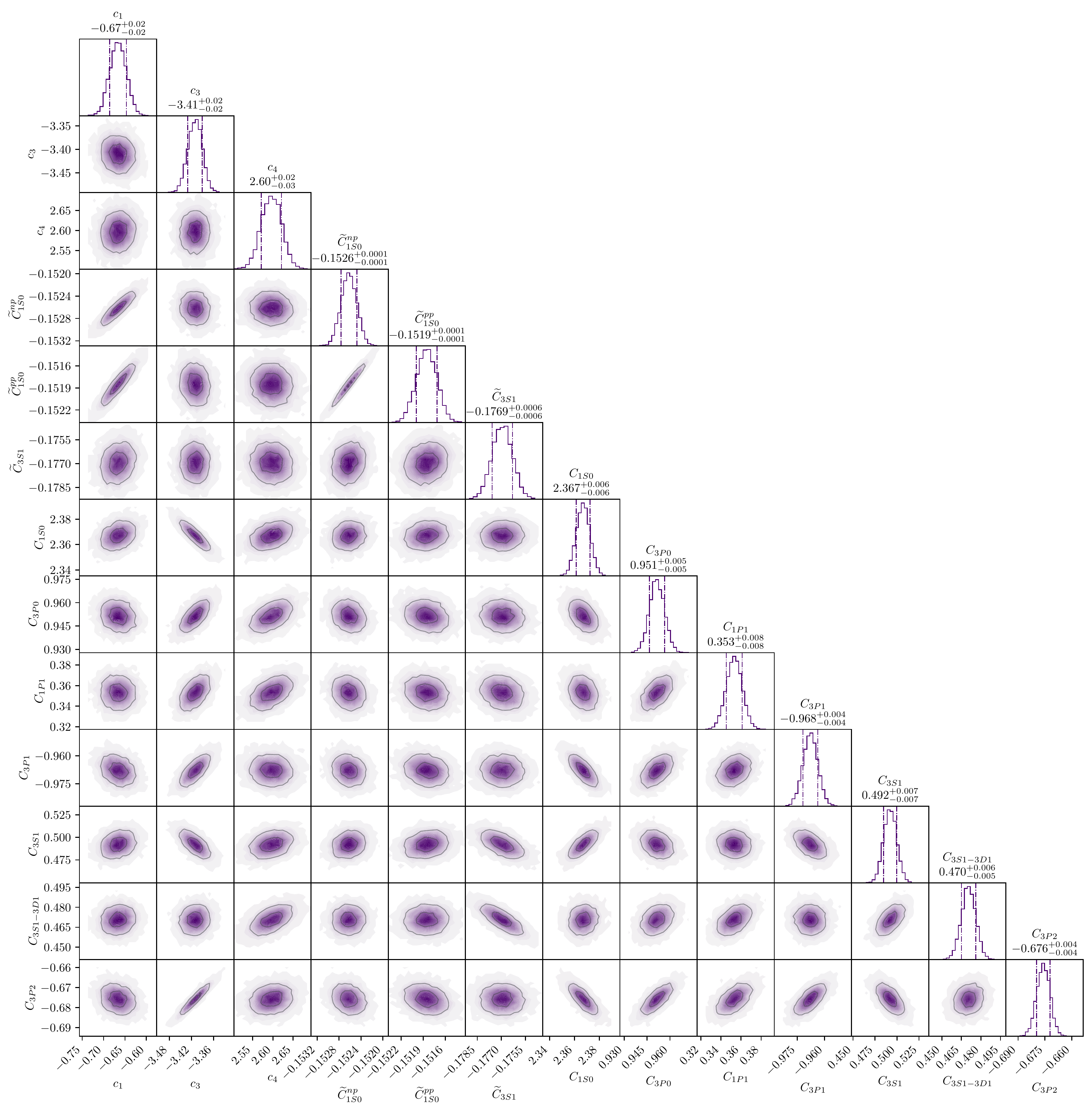}
\caption{\nnlo\ posterior sampled with HMC. The LECs are shown in units of $\Ctunit$ for the \lo\ contact LECs, $\Cunit$ for the \nlo\ contact LECs, and $\cunit$ for the \piN\ LECs. The inner (outer) gray contour line encloses 39\% (86\%) of the probability mass. The dot-dashed vertical lines indicate a 68\% credibility interval in the univariate marginals.}
\label{fig:\currfilebase}
\end{figure*}

%% file: texfig/n2lo_glob_piN_contours.tex
\begin{figure}[htpb]
\centering
\includegraphics[width=1.0\linewidth]{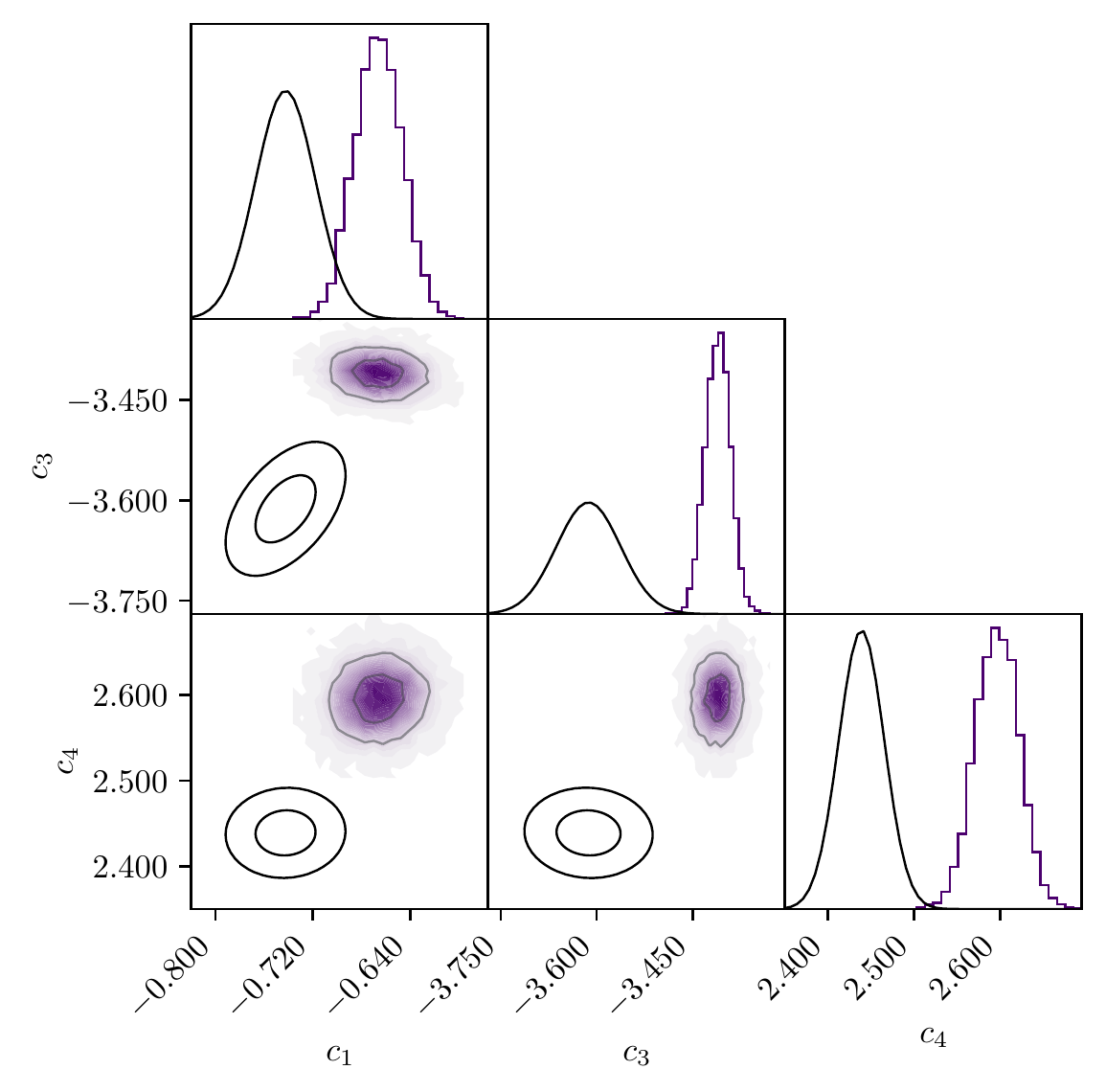}
\caption{Prior and posterior pdfs for the \piN{} LECs $c_1,c_3,c_4$, in units of GeV$^{-1}$, indicated with black-line ellipses and colored (purple jagged) regions, respectively. The inner (outer) black ellipses enclose 39\% (86\%) of the prior probability mass, and the jagged gray lines do the same for the posterior probability mass. The posteriors were obtained using HMC. See text and Fig.~\ref{fig:n2lo_glob_corner} for details.}
\label{fig:\currfilebase}
\end{figure}

%% file: texfig/gelman_rubin.tex
\begin{figure*}[htpb]
\centering
\includegraphics[width=1.0\linewidth]{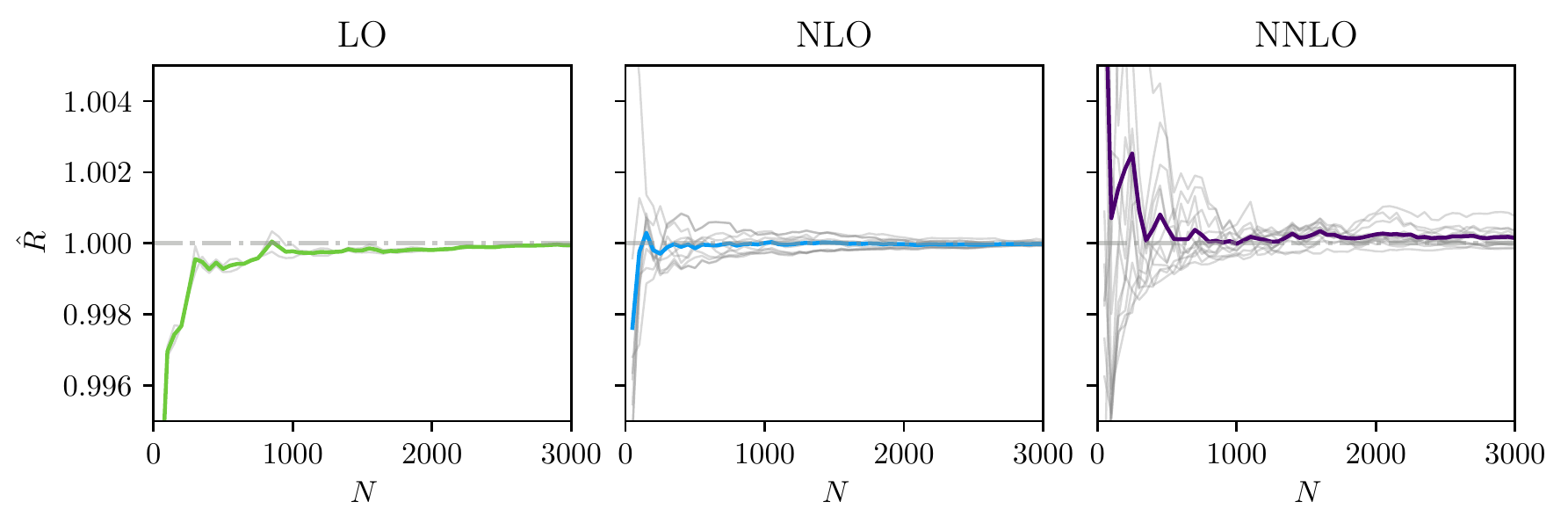}
\caption{The Gelman-Rubin convergence diagnostics $\hat{R}$ for the HMC sampled chains at \lo, \nlo, and \nnlo\ as a function of the number of samples $N$. $\hat{R}$ for each individual parameter is shown in gray, while the mean $\hat{R}$ is shown in green, blue, and purple, respectively. }
\label{fig:\currfilebase}
\end{figure*}

%% file: texfig/acors_comparison.tex
\begin{figure*}[htpb]
\centering
\includegraphics[width=1.0\linewidth]{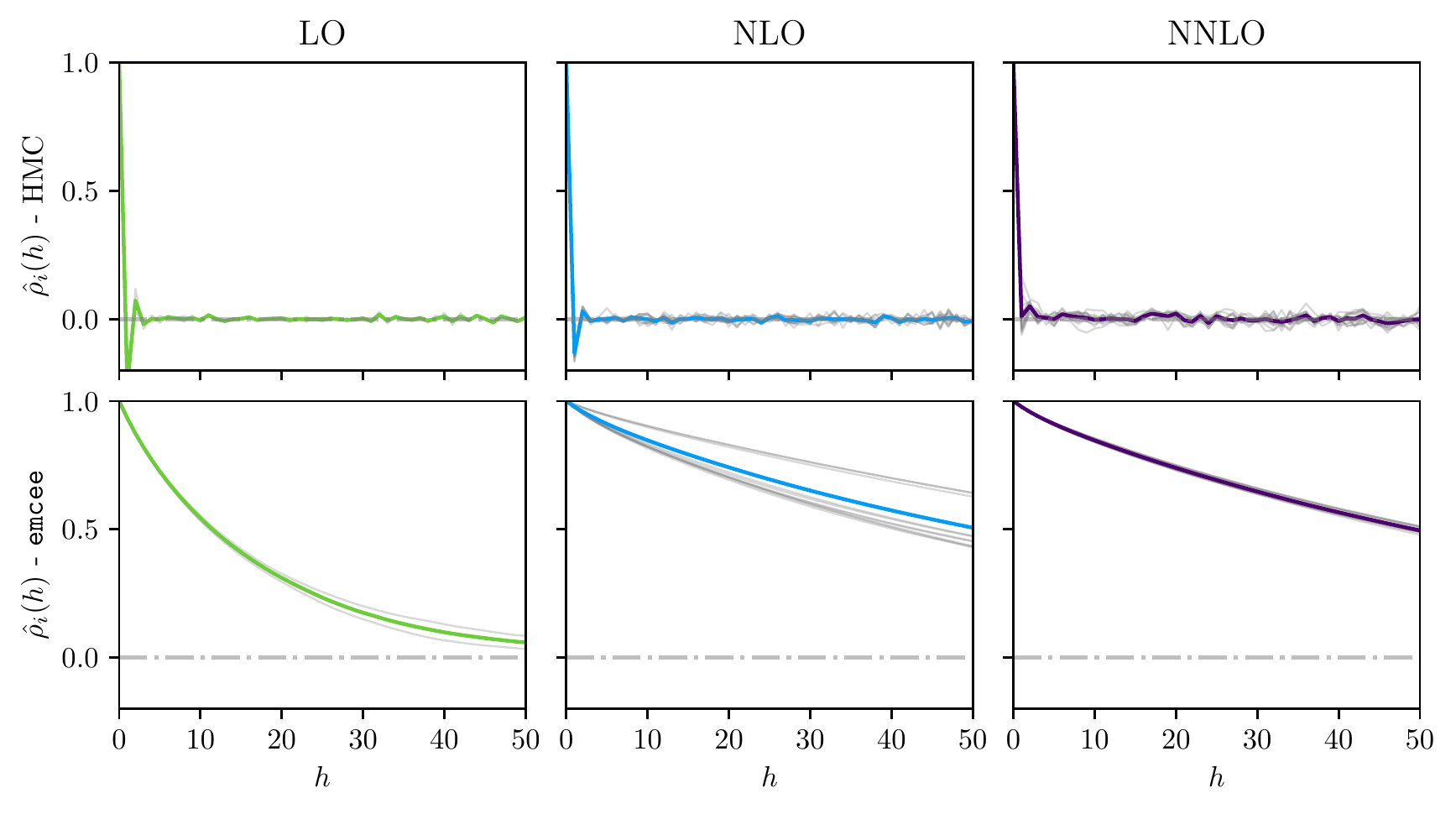}
\caption{Autocorrelation functions of the MCMC chains at LO-NNLO. The chains produced with HMC are shown in the first row and chains produced with \texttt{emcee} are shown in the second row. The gray lines show autocorrelations in individual parameters whereas the colored lines show the average over all parameters. The results are averaged over all walkers in the plots showing the \texttt{emcee} autocorrelations. See Table~\ref{tab:chain_stats} for details about each sampling.}
\label{fig:\currfilebase}
\end{figure*}

%% file: texfig/tau_vs_nsamples.tex
\begin{figure}[tb]
\centering
\includegraphics[width=1.0\linewidth]{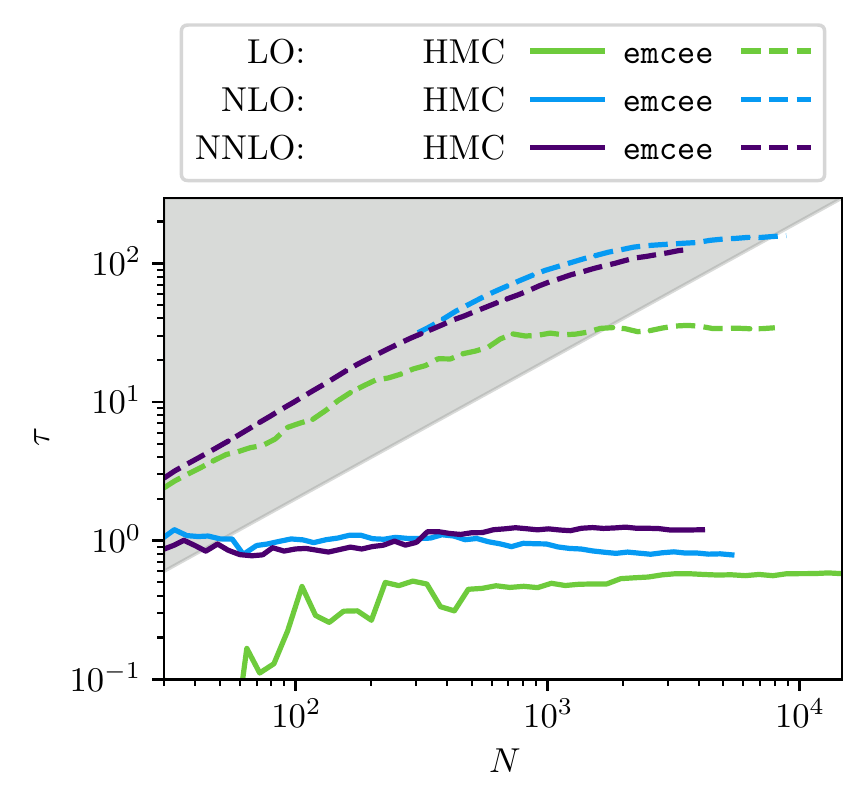}
\caption{Integrated autocorrelation time $\tau$ vs number of samples. The gray area indicates the zone of non-convergence, i.e., $N < 50\tau$. The \emcee\ results are averaged over all walkers, hence the seemingly low number of samples compared to Table~\ref{tab:ess}.}
\label{fig:\currfilebase}
\end{figure}

%% file: tables/ess.tex
\begin{table*}[tb]
\caption{\label{tab:\currfilebase} Comparison between the performance of \emcee\ and HMC applied to sample the LEC posteriors at \lo{}, \nlo{}, \nnlo{}. $\tau$ is the integrated autocorrelation time. ESS is the effective sample size and $N$ is the number of collected MCMC samples. Consequently, the ESS/$N$ column shows how many effective samples one nominal sample is worth. The \evalpersample\ column, where $N_\mathcal{L}$ is the total number of likelihood calls, shows the average number of likelihood evaluations (including tuning and burn-in) necessary to collect one nominal sample. $\mathcal{S}$ is the average real-world speedup of HMC compared to \emcee{} as defined in Eq.~\eqref{eq:speedup}. }
\begin{ruledtabular}
\begin{tabular}{c c c c c c c c c}
Algorithm & Order & $\tau$ & ESS/$N$ & $\evalpersample$ & $N_\mathcal{L}/$ESS & AD-cost & $\mathcal{S}$ \\
\colrule
\emcee & \lo   & 34     & 0.029       & 1.01 & 35     & \\
HMC    & \lo   & 0.58   & 1.7         & 9.07 & 5.3    & 1.10 & 6.0 \\
\emcee & \nlo  & 158    & 0.0063      & 1.07 & 169    & \\
HMC    & \nlo  & 0.78   & 1.3         & 27.5 & 21     & 1.24 & 6.4 \\
\emcee & \nnlo & $>126$ & $<0.007$    & 1.16 & $>146$ & \\
HMC    & \nnlo & 1.2    & 0.85        & 23.9 & 28     & 1.43 & $>3.6$ \\
\end{tabular}
\end{ruledtabular}
\end{table*}

%% file: texfig/nat_pub_acors.tex
\begin{figure}[tb]
\centering
\includegraphics[width=1.0\linewidth]{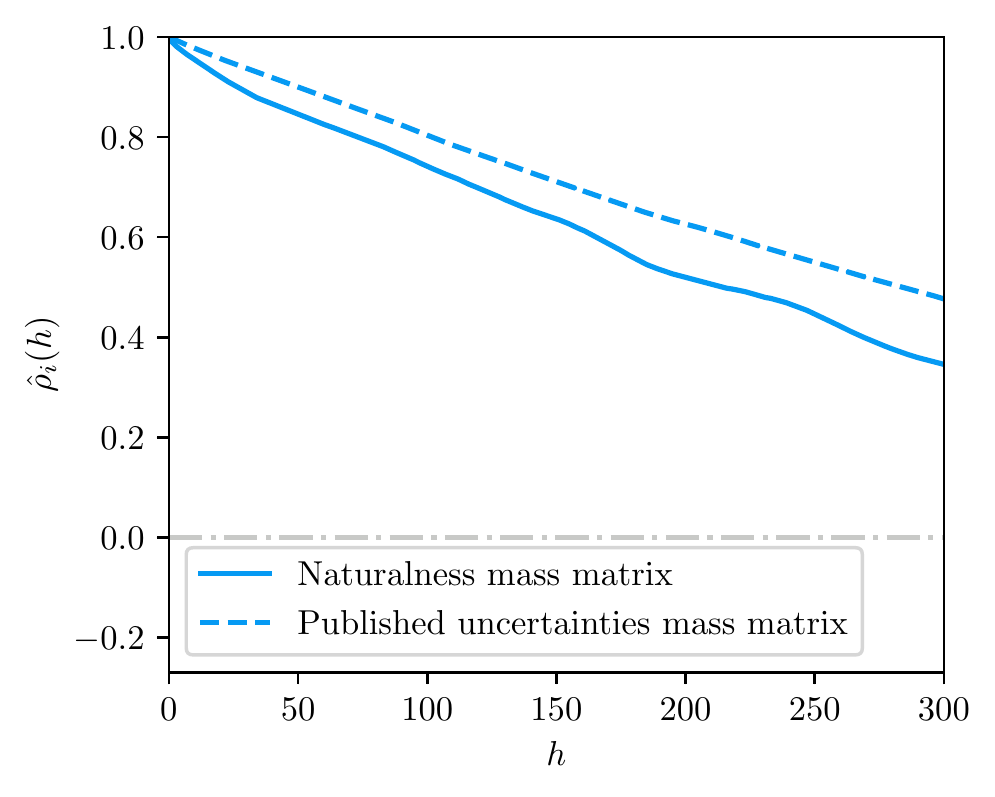}
\caption{Averaged autocorrelations at \nlo\ achieved with the ``naturalness'' and ``published uncertainties'' mass matrices.}
\label{fig:\currfilebase}
\end{figure}

%% file: texfig/lo_scan.tex
\begin{figure}[tp]
\centering
\includegraphics[width=1.0\linewidth]{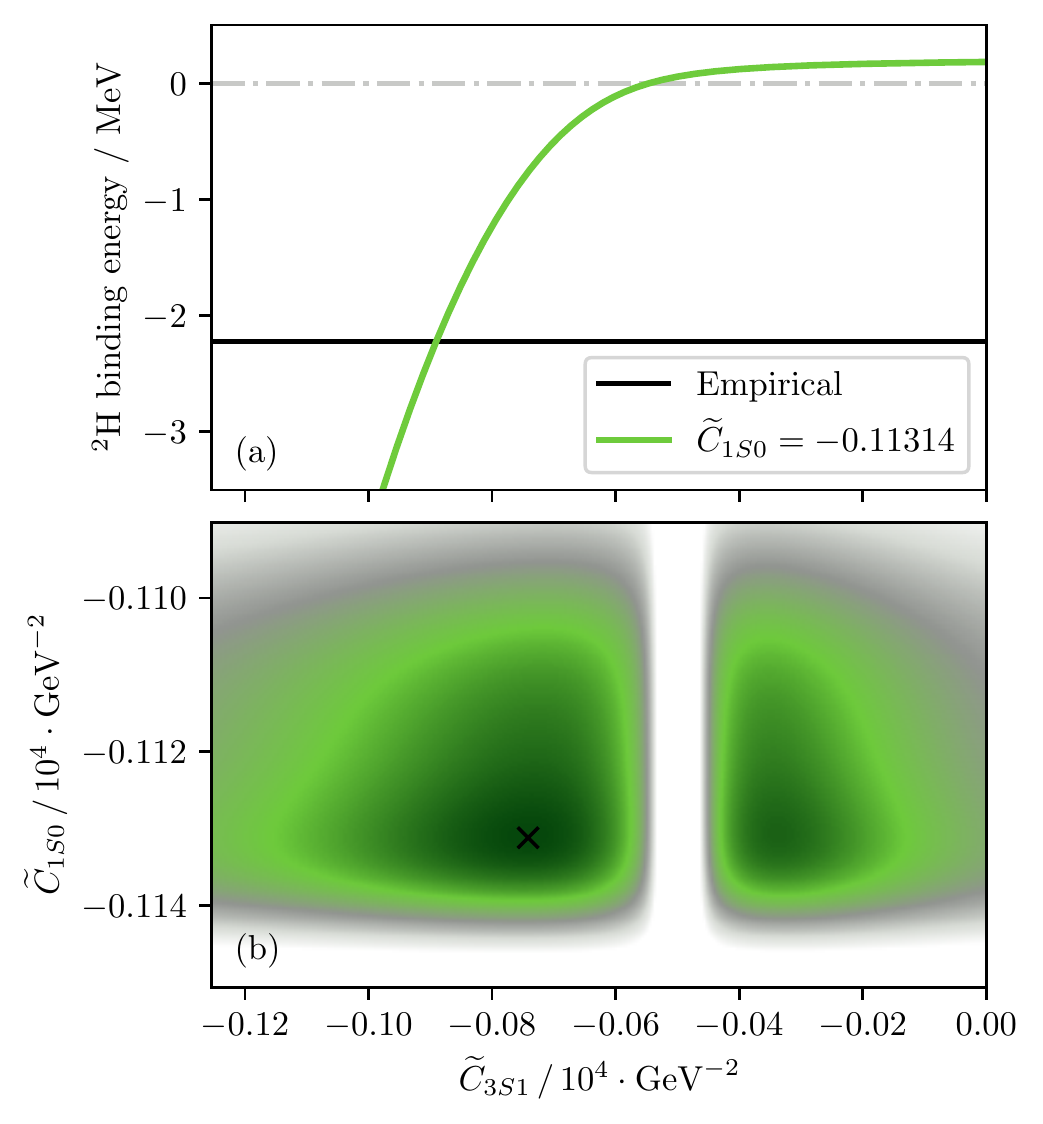}
\caption{(a) $^2$H binding energy as a function of $\Ct_{3S1}$ with $\Ct_{1S0} = -0.11314 \cdot \Ctunit$. (b) \lo\ posterior evaluated on a lattice. The black cross marks the MAP point indicated on the diagonal in Fig.~\ref{fig:lo_glob_corner}, i.e., $(\Ct_{1S0}, \Ct_{3S1}) = (-0.11312, -0.07416) \cdot \Ctunit$.}
\label{fig:\currfilebase}
\end{figure}

%% file: texfig/ppd_validation_385_215.tex
\begin{figure*}[htpb]
\centering
\includegraphics[width=1.0\linewidth]{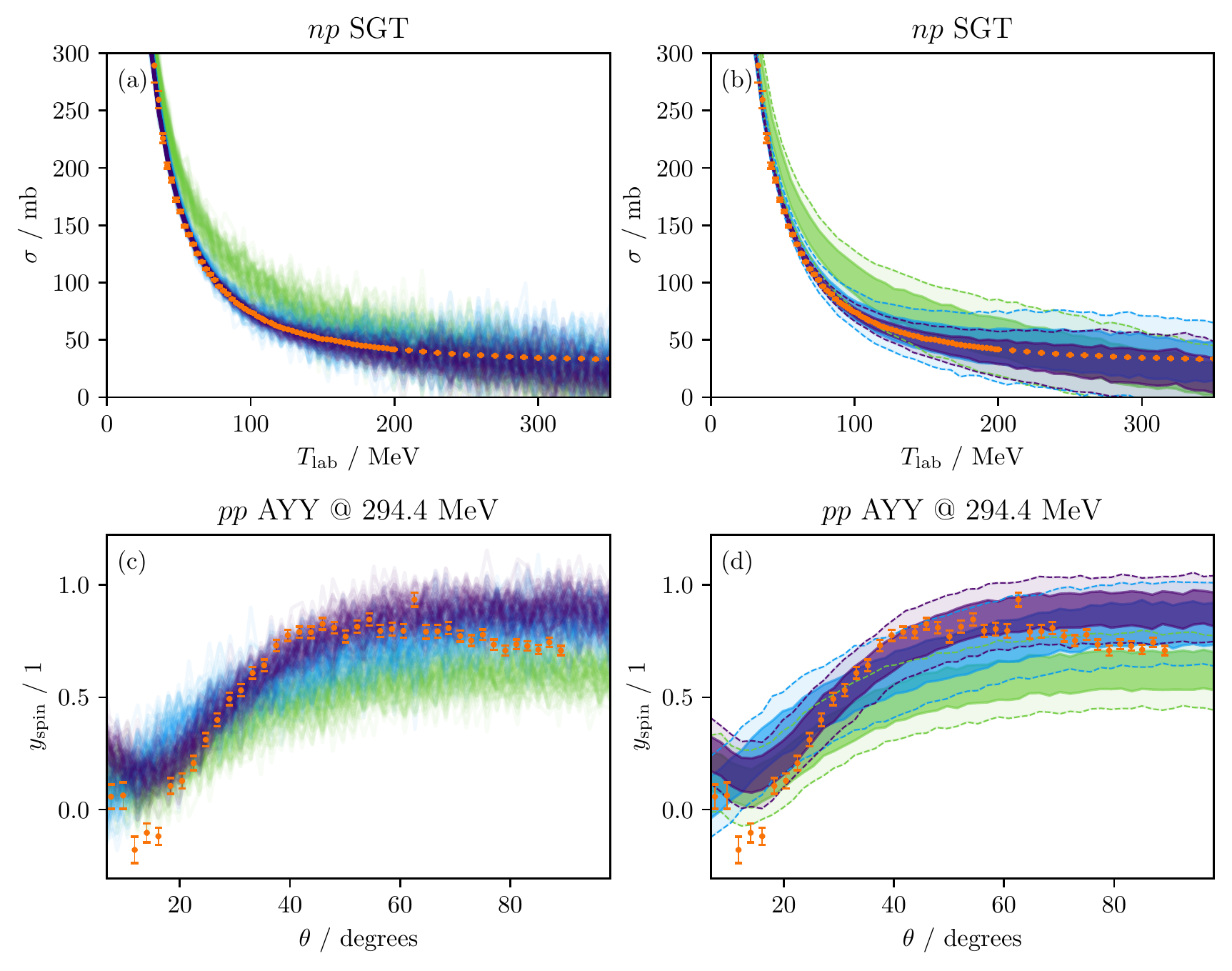}
\caption{Posterior predictive distributions for the true value of two \NN{} scattering observables at \lo{} (green), \nlo{} (blue), and \nnlo{} (purple). (a) Uncorrelated samples of $\prob(y_\text{true}|D,I)$ for the $np$ total cross section. Empirical data with error bars from Ref.~\cite{lisowski82} is shown in orange. (b) 68\% (dark shaded regions bounded by solid lines) and 95\% (light shaded regions bounded by dashed lines) HDIs of the ppds in (a). (c) Uncorrelated samples of $\prob(y_\text{true}|D,I)$ for $pp$ AYY at $T_\text{lab} = 294.4$ MeV. Empirical data with error bars from Ref.~\cite{vonPrzewoski:1998ye}. (d) 68 and 95\% HDIs of the ppds in (c).}
\label{fig:\currfilebase}
\end{figure*}

%% file: texfig/empirical_coverage.tex
\begin{figure*}[tb]
\centering
\includegraphics[width=1.0\linewidth]{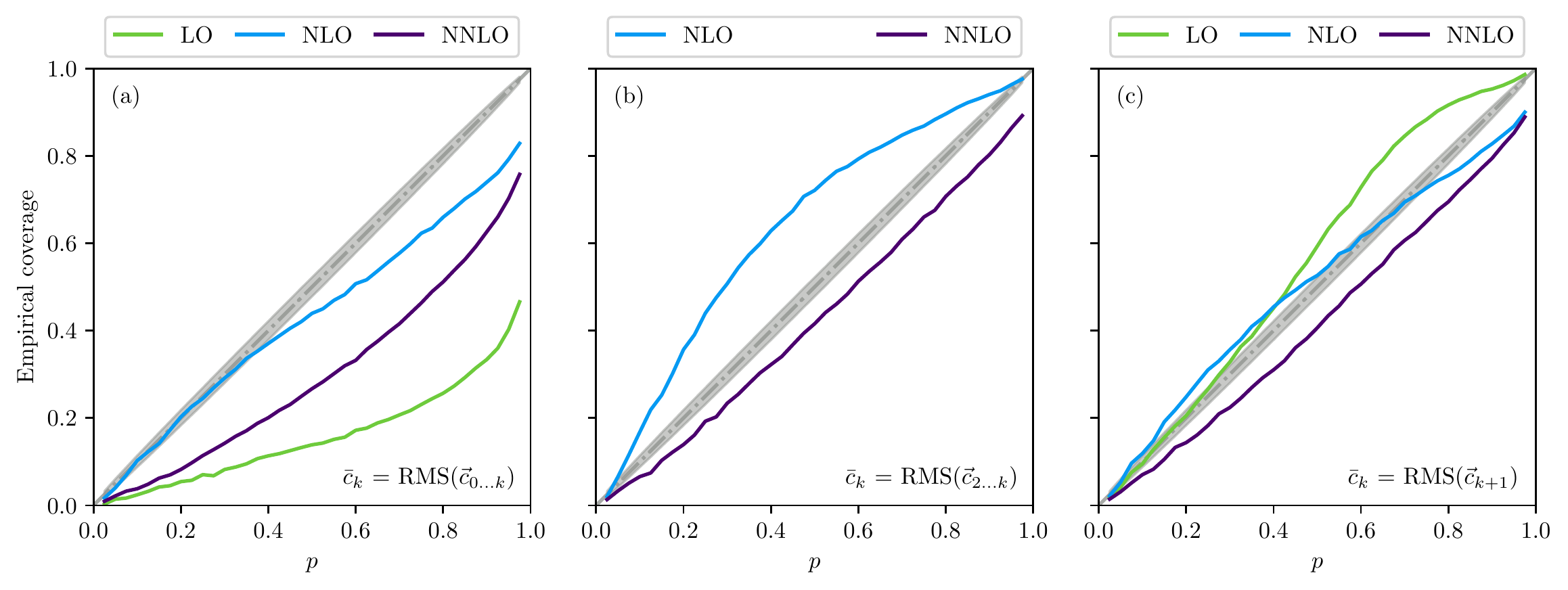}
\caption{Empirical coverage probabilities for $p \cdot 100$\% HDI credible intervals with respect to the validation data set $\Dval$. Each panel corresponds to a different method of computing $\cbar$. (a) $\cbar_k = \text{RMS}(\vec{c}_{0\ldots k})$, (b) $\cbar_k = \text{RMS}(\vec{c}_{2\ldots k})$, (c) $\cbar_k = \text{RMS}(\vec{c}_{k+1})$.
The computed $\cbar$ values are (a) $\cbar_{k=0} = 1.17$, $\cbar_{k=2} = 2.08$, $\cbar_{k=3} = 2.72$, (b) $\cbar_{k=2} = 4.95$, $\cbar_{k=3} = 4.19$, (c) $\cbar_{k=0} = 4.95$, $\cbar_{k=2} = 2.84$, $\cbar_{k=3} = 4.12$.}
\label{fig:\currfilebase}
\end{figure*}

%% file: texfig/empirical_coverage_subsets.tex
\begin{figure*}[htpb]
\centering
\includegraphics[width=1.0\linewidth]{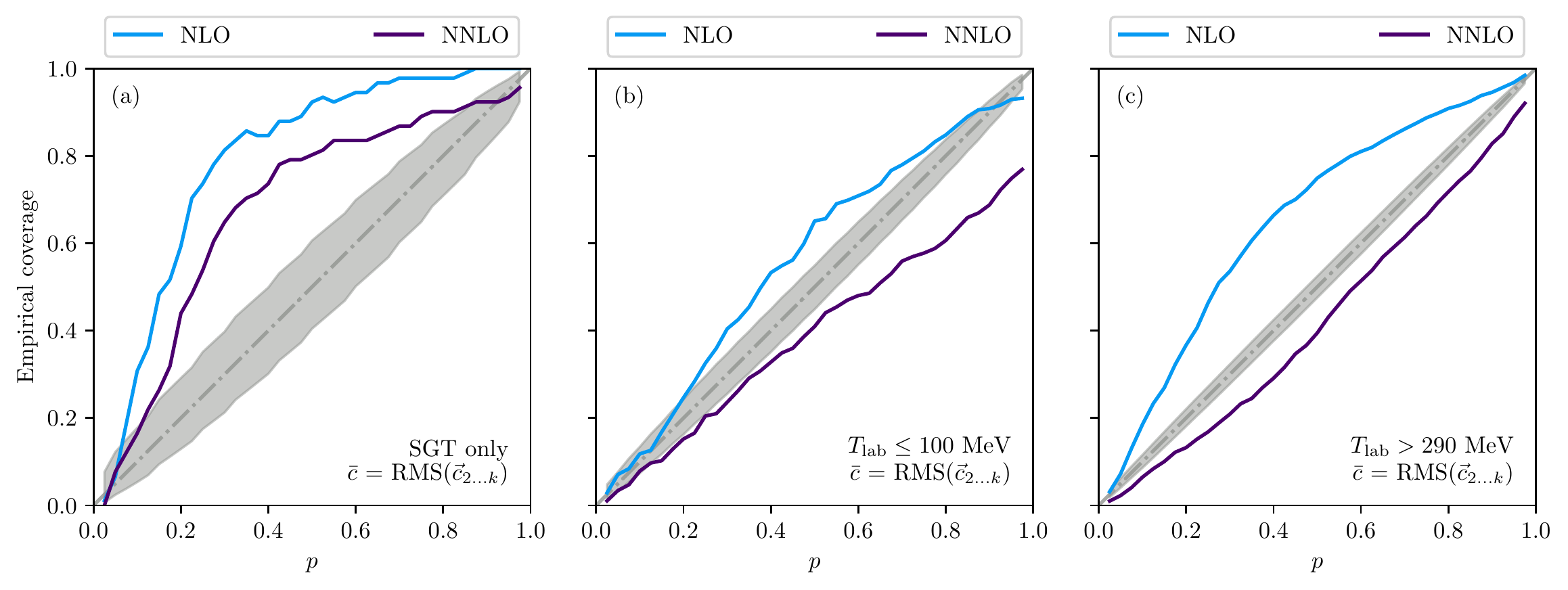}
\caption{Empirical coverage probabilities for $p \cdot 100$\% HDI credible intervals with respect to subsets of the validation data set $\Dval$. We compute $\cbar$ as $\cbar_k = \text{RMS}(c_{2\ldots k})$ as in Fig.~\ref{fig:empirical_coverage}(b) and use $\Lambda_b = 600$ MeV. (a) Only SGT observables, (b) all validation data with $\Tlab \leq 100$ MeV, (c) all validation data with $\Tlab > 290$ MeV.}
\label{fig:\currfilebase}
\end{figure*}

%% file: manuscript.bbl
\begin{thebibliography}{50}%
\makeatletter
\providecommand \@ifxundefined [1]{%
 \@ifx{#1\undefined}
}%
\providecommand \@ifnum [1]{%
 \ifnum #1\expandafter \@firstoftwo
 \else \expandafter \@secondoftwo
 \fi
}%
\providecommand \@ifx [1]{%
 \ifx #1\expandafter \@firstoftwo
 \else \expandafter \@secondoftwo
 \fi
}%
\providecommand \natexlab [1]{#1}%
\providecommand \enquote  [1]{``#1''}%
\providecommand \bibnamefont  [1]{#1}%
\providecommand \bibfnamefont [1]{#1}%
\providecommand \citenamefont [1]{#1}%
\providecommand \href@noop [0]{\@secondoftwo}%
\providecommand \href [0]{\begingroup \@sanitize@url \@href}%
\providecommand \@href[1]{\@@startlink{#1}\@@href}%
\providecommand \@@href[1]{\endgroup#1\@@endlink}%
\providecommand \@sanitize@url [0]{\catcode `\\12\catcode `\$12\catcode
  `\&12\catcode `\#12\catcode `\^12\catcode `\_12\catcode `\%12\relax}%
\providecommand \@@startlink[1]{}%
\providecommand \@@endlink[0]{}%
\providecommand \url  [0]{\begingroup\@sanitize@url \@url }%
\providecommand \@url [1]{\endgroup\@href {#1}{\urlprefix }}%
\providecommand \urlprefix  [0]{URL }%
\providecommand \Eprint [0]{\href }%
\providecommand \doibase [0]{http://dx.doi.org/}%
\providecommand \selectlanguage [0]{\@gobble}%
\providecommand \bibinfo  [0]{\@secondoftwo}%
\providecommand \bibfield  [0]{\@secondoftwo}%
\providecommand \translation [1]{[#1]}%
\providecommand \BibitemOpen [0]{}%
\providecommand \bibitemStop [0]{}%
\providecommand \bibitemNoStop [0]{.\EOS\space}%
\providecommand \EOS [0]{\spacefactor3000\relax}%
\providecommand \BibitemShut  [1]{\csname bibitem#1\endcsname}%
\let\auto@bib@innerbib\@empty
\bibitem [{\citenamefont {Bedaque}\ and\ \citenamefont {van
  Kolck}(2002)}]{Bedaque:2002mn}%
  \BibitemOpen
  \bibfield  {author} {\bibinfo {author} {\bibfnamefont {P.~F.}\ \bibnamefont
  {Bedaque}}\ and\ \bibinfo {author} {\bibfnamefont {U.}~\bibnamefont {van
  Kolck}},\ }\href {\doibase 10.1146/annurev.nucl.52.050102.090637} {\bibfield
  {journal} {\bibinfo  {journal} {Ann. Rev. Nucl. Part. Sci.}\ }\textbf
  {\bibinfo {volume} {52}},\ \bibinfo {pages} {339} (\bibinfo {year}
  {2002})}\BibitemShut {NoStop}%
\bibitem [{\citenamefont {Epelbaum}\ \emph {et~al.}(2009)\citenamefont
  {Epelbaum}, \citenamefont {Hammer},\ and\ \citenamefont
  {Meissner}}]{Epelbaum:2008ga}%
  \BibitemOpen
  \bibfield  {author} {\bibinfo {author} {\bibfnamefont {E.}~\bibnamefont
  {Epelbaum}}, \bibinfo {author} {\bibfnamefont {H.-W.}\ \bibnamefont
  {Hammer}}, \ and\ \bibinfo {author} {\bibfnamefont {U.-G.}\ \bibnamefont
  {Meissner}},\ }\href {\doibase 10.1103/RevModPhys.81.1773} {\bibfield
  {journal} {\bibinfo  {journal} {Rev. Mod. Phys.}\ }\textbf {\bibinfo {volume}
  {81}},\ \bibinfo {pages} {1773} (\bibinfo {year} {2009})}\BibitemShut
  {NoStop}%
\bibitem [{\citenamefont {Machleidt}\ and\ \citenamefont
  {Entem}(2011)}]{Machleidt:2011zz}%
  \BibitemOpen
  \bibfield  {author} {\bibinfo {author} {\bibfnamefont {R.}~\bibnamefont
  {Machleidt}}\ and\ \bibinfo {author} {\bibfnamefont {D.~R.}\ \bibnamefont
  {Entem}},\ }\href {\doibase 10.1016/j.physrep.2011.02.001} {\bibfield
  {journal} {\bibinfo  {journal} {Phys. Rept.}\ }\textbf {\bibinfo {volume}
  {503}},\ \bibinfo {pages} {1} (\bibinfo {year} {2011})}\BibitemShut {NoStop}%
\bibitem [{\citenamefont {Hammer}\ \emph {et~al.}(2020)\citenamefont {Hammer},
  \citenamefont {K\"onig},\ and\ \citenamefont {van Kolck}}]{Hammer:2019poc}%
  \BibitemOpen
  \bibfield  {author} {\bibinfo {author} {\bibfnamefont {H.~W.}\ \bibnamefont
  {Hammer}}, \bibinfo {author} {\bibfnamefont {S.}~\bibnamefont {K\"onig}}, \
  and\ \bibinfo {author} {\bibfnamefont {U.}~\bibnamefont {van Kolck}},\ }\href
  {\doibase 10.1103/RevModPhys.92.025004} {\bibfield  {journal} {\bibinfo
  {journal} {Rev. Mod. Phys.}\ }\textbf {\bibinfo {volume} {92}},\ \bibinfo
  {pages} {025004} (\bibinfo {year} {2020})}\BibitemShut {NoStop}%
\bibitem [{\citenamefont {Furnstahl}\ \emph {et~al.}(2015)\citenamefont
  {Furnstahl}, \citenamefont {Klco}, \citenamefont {Phillips},\ and\
  \citenamefont {Wesolowski}}]{furnstahl15}%
  \BibitemOpen
  \bibfield  {author} {\bibinfo {author} {\bibfnamefont {R.~J.}\ \bibnamefont
  {Furnstahl}}, \bibinfo {author} {\bibfnamefont {N.}~\bibnamefont {Klco}},
  \bibinfo {author} {\bibfnamefont {D.~R.}\ \bibnamefont {Phillips}}, \ and\
  \bibinfo {author} {\bibfnamefont {S.}~\bibnamefont {Wesolowski}},\ }\href
  {\doibase 10.1103/PhysRevC.92.024005} {\bibfield  {journal} {\bibinfo
  {journal} {Phys. Rev. C}\ }\textbf {\bibinfo {volume} {92}},\ \bibinfo
  {pages} {024005} (\bibinfo {year} {2015})}\BibitemShut {NoStop}%
\bibitem [{\citenamefont {Duane}\ \emph {et~al.}(1987)\citenamefont {Duane},
  \citenamefont {Kennedy}, \citenamefont {Pendleton},\ and\ \citenamefont
  {Roweth}}]{duane87}%
  \BibitemOpen
  \bibfield  {author} {\bibinfo {author} {\bibfnamefont {S.}~\bibnamefont
  {Duane}}, \bibinfo {author} {\bibfnamefont {A.}~\bibnamefont {Kennedy}},
  \bibinfo {author} {\bibfnamefont {B.~J.}\ \bibnamefont {Pendleton}}, \ and\
  \bibinfo {author} {\bibfnamefont {D.}~\bibnamefont {Roweth}},\ }\href
  {\doibase https://doi.org/10.1016/0370-2693(87)91197-X} {\bibfield  {journal}
  {\bibinfo  {journal} {Phys. Lett. B}\ }\textbf {\bibinfo {volume} {195}},\
  \bibinfo {pages} {216} (\bibinfo {year} {1987})}\BibitemShut {NoStop}%
\bibitem [{\citenamefont {Metropolis}\ \emph {et~al.}(1953)\citenamefont
  {Metropolis}, \citenamefont {Rosenbluth}, \citenamefont {Rosenbluth},
  \citenamefont {Teller},\ and\ \citenamefont {Teller}}]{metropolis53}%
  \BibitemOpen
  \bibfield  {author} {\bibinfo {author} {\bibfnamefont {N.}~\bibnamefont
  {Metropolis}}, \bibinfo {author} {\bibfnamefont {A.~W.}\ \bibnamefont
  {Rosenbluth}}, \bibinfo {author} {\bibfnamefont {M.~N.}\ \bibnamefont
  {Rosenbluth}}, \bibinfo {author} {\bibfnamefont {A.~H.}\ \bibnamefont
  {Teller}}, \ and\ \bibinfo {author} {\bibfnamefont {E.}~\bibnamefont
  {Teller}},\ }\href {\doibase 10.1063/1.1699114} {\bibfield  {journal}
  {\bibinfo  {journal} {J. Chem. Phys.}\ }\textbf {\bibinfo {volume} {21}},\
  \bibinfo {pages} {1087} (\bibinfo {year} {1953})}\BibitemShut {NoStop}%
\bibitem [{\citenamefont {Hastings}(1970)}]{Hastings:1970aa}%
  \BibitemOpen
  \bibfield  {author} {\bibinfo {author} {\bibfnamefont {W.~K.}\ \bibnamefont
  {Hastings}},\ }\href {\doibase 10.1093/biomet/57.1.97} {\bibfield  {journal}
  {\bibinfo  {journal} {Biometrika}\ }\textbf {\bibinfo {volume} {57}},\
  \bibinfo {pages} {97} (\bibinfo {year} {1970})}\BibitemShut {NoStop}%
\bibitem [{\citenamefont {Navarro~P\'erez}\ \emph {et~al.}(2013)\citenamefont
  {Navarro~P\'erez}, \citenamefont {Amaro},\ and\ \citenamefont
  {Ruiz~Arriola}}]{perez13-1}%
  \BibitemOpen
  \bibfield  {author} {\bibinfo {author} {\bibfnamefont {R.}~\bibnamefont
  {Navarro~P\'erez}}, \bibinfo {author} {\bibfnamefont {J.~E.}\ \bibnamefont
  {Amaro}}, \ and\ \bibinfo {author} {\bibfnamefont {E.}~\bibnamefont
  {Ruiz~Arriola}},\ }\href {\doibase 10.1103/PhysRevC.88.024002} {\bibfield
  {journal} {\bibinfo  {journal} {Phys. Rev. C}\ }\textbf {\bibinfo {volume}
  {88}},\ \bibinfo {pages} {024002} (\bibinfo {year} {2013})}\BibitemShut
  {NoStop}%
\bibitem [{\citenamefont {P\'erez}\ \emph {et~al.}(2013)\citenamefont
  {P\'erez}, \citenamefont {Amaro},\ and\ \citenamefont {Arriola}}]{perez13-2}%
  \BibitemOpen
  \bibfield  {author} {\bibinfo {author} {\bibfnamefont {R.~N.}\ \bibnamefont
  {P\'erez}}, \bibinfo {author} {\bibfnamefont {J.~E.}\ \bibnamefont {Amaro}},
  \ and\ \bibinfo {author} {\bibfnamefont {E.~R.}\ \bibnamefont {Arriola}},\
  }\href {\doibase 10.1103/PhysRevC.88.064002} {\bibfield  {journal} {\bibinfo
  {journal} {Phys. Rev. C}\ }\textbf {\bibinfo {volume} {88}},\ \bibinfo
  {pages} {064002} (\bibinfo {year} {2013})}\BibitemShut {NoStop}%
\bibitem [{\citenamefont {Wesolowski}\ \emph {et~al.}(2019)\citenamefont
  {Wesolowski}, \citenamefont {Furnstahl}, \citenamefont {Melendez},\ and\
  \citenamefont {Phillips}}]{wesolowski19}%
  \BibitemOpen
  \bibfield  {author} {\bibinfo {author} {\bibfnamefont {S.}~\bibnamefont
  {Wesolowski}}, \bibinfo {author} {\bibfnamefont {R.~J.}\ \bibnamefont
  {Furnstahl}}, \bibinfo {author} {\bibfnamefont {J.~A.}\ \bibnamefont
  {Melendez}}, \ and\ \bibinfo {author} {\bibfnamefont {D.~R.}\ \bibnamefont
  {Phillips}},\ }\href {\doibase 10.1088/1361-6471/aaf5fc} {\bibfield
  {journal} {\bibinfo  {journal} {J. Phys. G}\ }\textbf {\bibinfo {volume}
  {46}},\ \bibinfo {pages} {045102} (\bibinfo {year} {2019})}\BibitemShut
  {NoStop}%
\bibitem [{\citenamefont {Hoferichter}\ \emph {et~al.}(2016)\citenamefont
  {Hoferichter}, \citenamefont {Ruiz~de Elvira}, \citenamefont {Kubis},\ and\
  \citenamefont {Mei\ss{}ner}}]{Hoferichter:2015hva}%
  \BibitemOpen
  \bibfield  {author} {\bibinfo {author} {\bibfnamefont {M.}~\bibnamefont
  {Hoferichter}}, \bibinfo {author} {\bibfnamefont {J.}~\bibnamefont {Ruiz~de
  Elvira}}, \bibinfo {author} {\bibfnamefont {B.}~\bibnamefont {Kubis}}, \ and\
  \bibinfo {author} {\bibfnamefont {U.-G.}\ \bibnamefont {Mei\ss{}ner}},\
  }\href {\doibase 10.1016/j.physrep.2016.02.002} {\bibfield  {journal}
  {\bibinfo  {journal} {Phys. Rept.}\ }\textbf {\bibinfo {volume} {625}},\
  \bibinfo {pages} {1} (\bibinfo {year} {2016})}\BibitemShut {NoStop}%
\bibitem [{\citenamefont {Siemens}\ \emph {et~al.}(2017)\citenamefont
  {Siemens}, \citenamefont {{Ruiz de Elvira}}, \citenamefont {Epelbaum},
  \citenamefont {Hoferichter}, \citenamefont {Krebs}, \citenamefont {Kubis},\
  and\ \citenamefont {Meißner}}]{siemens17}%
  \BibitemOpen
  \bibfield  {author} {\bibinfo {author} {\bibfnamefont {D.}~\bibnamefont
  {Siemens}}, \bibinfo {author} {\bibfnamefont {J.}~\bibnamefont {{Ruiz de
  Elvira}}}, \bibinfo {author} {\bibfnamefont {E.}~\bibnamefont {Epelbaum}},
  \bibinfo {author} {\bibfnamefont {M.}~\bibnamefont {Hoferichter}}, \bibinfo
  {author} {\bibfnamefont {H.}~\bibnamefont {Krebs}}, \bibinfo {author}
  {\bibfnamefont {B.}~\bibnamefont {Kubis}}, \ and\ \bibinfo {author}
  {\bibfnamefont {U.}~\bibnamefont {Meißner}},\ }\href {\doibase
  https://doi.org/10.1016/j.physletb.2017.04.039} {\bibfield  {journal}
  {\bibinfo  {journal} {Phys. Lett. B}\ }\textbf {\bibinfo {volume} {770}},\
  \bibinfo {pages} {27 } (\bibinfo {year} {2017})}\BibitemShut {NoStop}%
\bibitem [{\citenamefont {Carlsson}\ \emph {et~al.}(2016)\citenamefont
  {Carlsson}, \citenamefont {Ekstr\"om}, \citenamefont {Forss\'en},
  \citenamefont {Str\"omberg}, \citenamefont {Jansen}, \citenamefont {Lilja},
  \citenamefont {Lindby}, \citenamefont {Mattsson},\ and\ \citenamefont
  {Wendt}}]{Carlsson:2015vda}%
  \BibitemOpen
  \bibfield  {author} {\bibinfo {author} {\bibfnamefont {B.~D.}\ \bibnamefont
  {Carlsson}}, \bibinfo {author} {\bibfnamefont {A.}~\bibnamefont {Ekstr\"om}},
  \bibinfo {author} {\bibfnamefont {C.}~\bibnamefont {Forss\'en}}, \bibinfo
  {author} {\bibfnamefont {D.~F.}\ \bibnamefont {Str\"omberg}}, \bibinfo
  {author} {\bibfnamefont {G.~R.}\ \bibnamefont {Jansen}}, \bibinfo {author}
  {\bibfnamefont {O.}~\bibnamefont {Lilja}}, \bibinfo {author} {\bibfnamefont
  {M.}~\bibnamefont {Lindby}}, \bibinfo {author} {\bibfnamefont {B.~A.}\
  \bibnamefont {Mattsson}}, \ and\ \bibinfo {author} {\bibfnamefont {K.~A.}\
  \bibnamefont {Wendt}},\ }\href {\doibase 10.1103/PhysRevX.6.011019}
  {\bibfield  {journal} {\bibinfo  {journal} {Phys. Rev. X}\ }\textbf {\bibinfo
  {volume} {6}},\ \bibinfo {pages} {011019} (\bibinfo {year}
  {2016})}\BibitemShut {NoStop}%
\bibitem [{\citenamefont {Wesolowski}\ \emph {et~al.}(2016)\citenamefont
  {Wesolowski}, \citenamefont {Klco}, \citenamefont {Furnstahl}, \citenamefont
  {Phillips},\ and\ \citenamefont {Thapaliya}}]{wesolowski16}%
  \BibitemOpen
  \bibfield  {author} {\bibinfo {author} {\bibfnamefont {S.}~\bibnamefont
  {Wesolowski}}, \bibinfo {author} {\bibfnamefont {N.}~\bibnamefont {Klco}},
  \bibinfo {author} {\bibfnamefont {R.~J.}\ \bibnamefont {Furnstahl}}, \bibinfo
  {author} {\bibfnamefont {D.~R.}\ \bibnamefont {Phillips}}, \ and\ \bibinfo
  {author} {\bibfnamefont {A.}~\bibnamefont {Thapaliya}},\ }\href {\doibase
  10.1088/0954-3899/43/7/074001} {\bibfield  {journal} {\bibinfo  {journal} {J.
  Phys. G}\ }\textbf {\bibinfo {volume} {43}},\ \bibinfo {pages} {074001}
  (\bibinfo {year} {2016})}\BibitemShut {NoStop}%
\bibitem [{\citenamefont {Lisowski}\ \emph {et~al.}(1982)\citenamefont
  {Lisowski}, \citenamefont {Shamu}, \citenamefont {Auchampaugh}, \citenamefont
  {King}, \citenamefont {Moore}, \citenamefont {Morgan},\ and\ \citenamefont
  {Singleton}}]{lisowski82}%
  \BibitemOpen
  \bibfield  {author} {\bibinfo {author} {\bibfnamefont {P.~W.}\ \bibnamefont
  {Lisowski}}, \bibinfo {author} {\bibfnamefont {R.~E.}\ \bibnamefont {Shamu}},
  \bibinfo {author} {\bibfnamefont {G.~F.}\ \bibnamefont {Auchampaugh}},
  \bibinfo {author} {\bibfnamefont {N.~S.~P.}\ \bibnamefont {King}}, \bibinfo
  {author} {\bibfnamefont {M.~S.}\ \bibnamefont {Moore}}, \bibinfo {author}
  {\bibfnamefont {G.~L.}\ \bibnamefont {Morgan}}, \ and\ \bibinfo {author}
  {\bibfnamefont {T.~S.}\ \bibnamefont {Singleton}},\ }\href {\doibase
  10.1103/PhysRevLett.49.255} {\bibfield  {journal} {\bibinfo  {journal} {Phys.
  Rev. Lett.}\ }\textbf {\bibinfo {volume} {49}},\ \bibinfo {pages} {255}
  (\bibinfo {year} {1982})}\BibitemShut {NoStop}%
\bibitem [{SAI()}]{SAID}%
  \BibitemOpen
  \href@noop {} {\enquote {\bibinfo {title} {Ins data analysis center:
  {SAID}},}\ }\bibinfo {howpublished} {http://gwdac.phys.gwu.edu},\ \bibinfo
  {note} {accessed: 2021-03-29}\BibitemShut {NoStop}%
\bibitem [{\citenamefont {Bergervoet}\ \emph {et~al.}(1988)\citenamefont
  {Bergervoet}, \citenamefont {van Campen}, \citenamefont {van~der Sanden},\
  and\ \citenamefont {de~Swart}}]{Bergervoet:1988zz}%
  \BibitemOpen
  \bibfield  {author} {\bibinfo {author} {\bibfnamefont {J.~R.}\ \bibnamefont
  {Bergervoet}}, \bibinfo {author} {\bibfnamefont {P.~C.}\ \bibnamefont {van
  Campen}}, \bibinfo {author} {\bibfnamefont {W.~A.}\ \bibnamefont {van~der
  Sanden}}, \ and\ \bibinfo {author} {\bibfnamefont {J.~J.}\ \bibnamefont
  {de~Swart}},\ }\href {\doibase 10.1103/PhysRevC.38.15} {\bibfield  {journal}
  {\bibinfo  {journal} {Phys. Rev. C}\ }\textbf {\bibinfo {volume} {38}},\
  \bibinfo {pages} {15} (\bibinfo {year} {1988})}\BibitemShut {NoStop}%
\bibitem [{\citenamefont {Bergervoet}\ \emph {et~al.}(1990)\citenamefont
  {Bergervoet}, \citenamefont {van Campen}, \citenamefont {Klomp},
  \citenamefont {de~Kok}, \citenamefont {Rijken}, \citenamefont {Stoks},\ and\
  \citenamefont {de~Swart}}]{Bergervoet:1990zy}%
  \BibitemOpen
  \bibfield  {author} {\bibinfo {author} {\bibfnamefont {J.~R.}\ \bibnamefont
  {Bergervoet}}, \bibinfo {author} {\bibfnamefont {P.~C.}\ \bibnamefont {van
  Campen}}, \bibinfo {author} {\bibfnamefont {R.~A.~M.}\ \bibnamefont {Klomp}},
  \bibinfo {author} {\bibfnamefont {J.~L.}\ \bibnamefont {de~Kok}}, \bibinfo
  {author} {\bibfnamefont {T.~A.}\ \bibnamefont {Rijken}}, \bibinfo {author}
  {\bibfnamefont {V.~G.~J.}\ \bibnamefont {Stoks}}, \ and\ \bibinfo {author}
  {\bibfnamefont {J.~J.}\ \bibnamefont {de~Swart}},\ }\href {\doibase
  10.1103/PhysRevC.41.1435} {\bibfield  {journal} {\bibinfo  {journal} {Phys.
  Rev. C}\ }\textbf {\bibinfo {volume} {41}},\ \bibinfo {pages} {1435}
  (\bibinfo {year} {1990})}\BibitemShut {NoStop}%
\bibitem [{\citenamefont {Stump}\ \emph {et~al.}(2001)\citenamefont {Stump},
  \citenamefont {Pumplin}, \citenamefont {Brock}, \citenamefont {Casey},
  \citenamefont {Huston}, \citenamefont {Kalk}, \citenamefont {Lai},\ and\
  \citenamefont {Tung}}]{Stump:2001gu}%
  \BibitemOpen
  \bibfield  {author} {\bibinfo {author} {\bibfnamefont {D.}~\bibnamefont
  {Stump}}, \bibinfo {author} {\bibfnamefont {J.}~\bibnamefont {Pumplin}},
  \bibinfo {author} {\bibfnamefont {R.}~\bibnamefont {Brock}}, \bibinfo
  {author} {\bibfnamefont {D.}~\bibnamefont {Casey}}, \bibinfo {author}
  {\bibfnamefont {J.}~\bibnamefont {Huston}}, \bibinfo {author} {\bibfnamefont
  {J.}~\bibnamefont {Kalk}}, \bibinfo {author} {\bibfnamefont {H.~L.}\
  \bibnamefont {Lai}}, \ and\ \bibinfo {author} {\bibfnamefont {W.~K.}\
  \bibnamefont {Tung}},\ }\href {\doibase 10.1103/PhysRevD.65.014012}
  {\bibfield  {journal} {\bibinfo  {journal} {Phys. Rev. D}\ }\textbf {\bibinfo
  {volume} {65}},\ \bibinfo {pages} {014012} (\bibinfo {year}
  {2001})}\BibitemShut {NoStop}%
\bibitem [{\citenamefont {Melendez}\ \emph {et~al.}(2019)\citenamefont
  {Melendez}, \citenamefont {Furnstahl}, \citenamefont {Phillips},
  \citenamefont {Pratola},\ and\ \citenamefont
  {Wesolowski}}]{Melendez:2019izc}%
  \BibitemOpen
  \bibfield  {author} {\bibinfo {author} {\bibfnamefont {J.~A.}\ \bibnamefont
  {Melendez}}, \bibinfo {author} {\bibfnamefont {R.~J.}\ \bibnamefont
  {Furnstahl}}, \bibinfo {author} {\bibfnamefont {D.~R.}\ \bibnamefont
  {Phillips}}, \bibinfo {author} {\bibfnamefont {M.~T.}\ \bibnamefont
  {Pratola}}, \ and\ \bibinfo {author} {\bibfnamefont {S.}~\bibnamefont
  {Wesolowski}},\ }\href {\doibase 10.1103/PhysRevC.100.044001} {\bibfield
  {journal} {\bibinfo  {journal} {Phys. Rev. C}\ }\textbf {\bibinfo {volume}
  {100}},\ \bibinfo {pages} {044001} (\bibinfo {year} {2019})}\BibitemShut
  {NoStop}%
\bibitem [{\citenamefont {Neal}(2011)}]{neal11}%
  \BibitemOpen
  \bibfield  {author} {\bibinfo {author} {\bibfnamefont {R.~M.}\ \bibnamefont
  {Neal}},\ }in\ \href@noop {} {\emph {\bibinfo {booktitle} {Handbook of Markov
  Chain Monte Carlo}}},\ \bibinfo {editor} {edited by\ \bibinfo {editor}
  {\bibfnamefont {S.}~\bibnamefont {Brooks}}, \bibinfo {editor} {\bibfnamefont
  {A.}~\bibnamefont {Gelman}}, \bibinfo {editor} {\bibfnamefont {G.~L.}\
  \bibnamefont {Jones}}, \ and\ \bibinfo {editor} {\bibfnamefont {X.-L.}\
  \bibnamefont {Meng}}}\ (\bibinfo  {publisher} {CRC Press},\ \bibinfo
  {address} {Boca Raton},\ \bibinfo {year} {2011})\ pp.\ \bibinfo {pages} {113
  -- 162}\BibitemShut {NoStop}%
\bibitem [{\citenamefont {Van~Rossum}\ and\ \citenamefont
  {Drake}(2009)}]{van_rossum09}%
  \BibitemOpen
  \bibfield  {author} {\bibinfo {author} {\bibfnamefont {G.}~\bibnamefont
  {Van~Rossum}}\ and\ \bibinfo {author} {\bibfnamefont {F.~L.}\ \bibnamefont
  {Drake}},\ }\href@noop {} {\emph {\bibinfo {title} {Python 3 Reference
  Manual}}}\ (\bibinfo  {publisher} {CreateSpace},\ \bibinfo {address} {Scotts
  Valley, CA},\ \bibinfo {year} {2009})\BibitemShut {NoStop}%
\bibitem [{\citenamefont {Harris}\ \emph {et~al.}(2020)\citenamefont {Harris},
  \citenamefont {Millman}, \citenamefont {van~der Walt}, \citenamefont
  {Gommers}, \citenamefont {Virtanen}, \citenamefont {Cournapeau},
  \citenamefont {Wieser}, \citenamefont {Taylor}, \citenamefont {Berg},
  \citenamefont {Smith}, \citenamefont {Kern}, \citenamefont {Picus},
  \citenamefont {Hoyer}, \citenamefont {van Kerkwijk}, \citenamefont {Brett},
  \citenamefont {Haldane}, \citenamefont {Fernández~del Río}, \citenamefont
  {Wiebe}, \citenamefont {Peterson}, \citenamefont {Gérard-Marchant},
  \citenamefont {Sheppard}, \citenamefont {Reddy}, \citenamefont {Weckesser},
  \citenamefont {Abbasi}, \citenamefont {Gohlke},\ and\ \citenamefont
  {Oliphant}}]{harris20}%
  \BibitemOpen
  \bibfield  {author} {\bibinfo {author} {\bibfnamefont {C.~R.}\ \bibnamefont
  {Harris}}, \bibinfo {author} {\bibfnamefont {K.~J.}\ \bibnamefont {Millman}},
  \bibinfo {author} {\bibfnamefont {S.~J.}\ \bibnamefont {van~der Walt}},
  \bibinfo {author} {\bibfnamefont {R.}~\bibnamefont {Gommers}}, \bibinfo
  {author} {\bibfnamefont {P.}~\bibnamefont {Virtanen}}, \bibinfo {author}
  {\bibfnamefont {D.}~\bibnamefont {Cournapeau}}, \bibinfo {author}
  {\bibfnamefont {E.}~\bibnamefont {Wieser}}, \bibinfo {author} {\bibfnamefont
  {J.}~\bibnamefont {Taylor}}, \bibinfo {author} {\bibfnamefont
  {S.}~\bibnamefont {Berg}}, \bibinfo {author} {\bibfnamefont {N.~J.}\
  \bibnamefont {Smith}}, \bibinfo {author} {\bibfnamefont {R.}~\bibnamefont
  {Kern}}, \bibinfo {author} {\bibfnamefont {M.}~\bibnamefont {Picus}},
  \bibinfo {author} {\bibfnamefont {S.}~\bibnamefont {Hoyer}}, \bibinfo
  {author} {\bibfnamefont {M.~H.}\ \bibnamefont {van Kerkwijk}}, \bibinfo
  {author} {\bibfnamefont {M.}~\bibnamefont {Brett}}, \bibinfo {author}
  {\bibfnamefont {A.}~\bibnamefont {Haldane}}, \bibinfo {author} {\bibfnamefont
  {J.}~\bibnamefont {Fernández~del Río}}, \bibinfo {author} {\bibfnamefont
  {M.}~\bibnamefont {Wiebe}}, \bibinfo {author} {\bibfnamefont
  {P.}~\bibnamefont {Peterson}}, \bibinfo {author} {\bibfnamefont
  {P.}~\bibnamefont {Gérard-Marchant}}, \bibinfo {author} {\bibfnamefont
  {K.}~\bibnamefont {Sheppard}}, \bibinfo {author} {\bibfnamefont
  {T.}~\bibnamefont {Reddy}}, \bibinfo {author} {\bibfnamefont
  {W.}~\bibnamefont {Weckesser}}, \bibinfo {author} {\bibfnamefont
  {H.}~\bibnamefont {Abbasi}}, \bibinfo {author} {\bibfnamefont
  {C.}~\bibnamefont {Gohlke}}, \ and\ \bibinfo {author} {\bibfnamefont {T.~E.}\
  \bibnamefont {Oliphant}},\ }\href {\doibase 10.1038/s41586-020-2649-2}
  {\bibfield  {journal} {\bibinfo  {journal} {Nature}\ }\textbf {\bibinfo
  {volume} {585}},\ \bibinfo {pages} {357} (\bibinfo {year}
  {2020})}\BibitemShut {NoStop}%
\bibitem [{\citenamefont {{Stan Development Team}}(2021)}]{stan_manual21}%
  \BibitemOpen
  \bibfield  {author} {\bibinfo {author} {\bibnamefont {{Stan Development
  Team}}},\ }\href {https://mc-stan.org} {\enquote {\bibinfo {title} {Stan
  modeling language users guide and reference manual, version 2.26},}\ }
  (\bibinfo {year} {2021})\BibitemShut {NoStop}%
\bibitem [{\citenamefont {Betancourt}()}]{betancourt18}%
  \BibitemOpen
  \bibfield  {author} {\bibinfo {author} {\bibfnamefont {M.}~\bibnamefont
  {Betancourt}},\ }\href@noop {} {}\Eprint {http://arxiv.org/abs/1701.02434}
  {arXiv:1701.02434} \BibitemShut {NoStop}%
\bibitem [{\citenamefont {Griewank}(2003)}]{Griewank:2003}%
  \BibitemOpen
  \bibfield  {author} {\bibinfo {author} {\bibfnamefont {A.}~\bibnamefont
  {Griewank}},\ }\href {\doibase 10.1017/S0962492902000132} {\bibfield
  {journal} {\bibinfo  {journal} {Acta Numer.}\ }\textbf {\bibinfo {volume}
  {12}},\ \bibinfo {pages} {321} (\bibinfo {year} {2003})}\BibitemShut
  {NoStop}%
\bibitem [{\citenamefont {Charpentier}\ and\ \citenamefont
  {Utke}(2009)}]{charpentier09}%
  \BibitemOpen
  \bibfield  {author} {\bibinfo {author} {\bibfnamefont {I.}~\bibnamefont
  {Charpentier}}\ and\ \bibinfo {author} {\bibfnamefont {J.}~\bibnamefont
  {Utke}},\ }\href {\doibase 10.1080/10556780802413769} {\bibfield  {journal}
  {\bibinfo  {journal} {Optim. Method. Softw.}\ }\textbf {\bibinfo {volume}
  {24}},\ \bibinfo {pages} {1} (\bibinfo {year} {2009})}\BibitemShut {NoStop}%
\bibitem [{\citenamefont {Homan}\ and\ \citenamefont {Gelman}(2014)}]{homan14}%
  \BibitemOpen
  \bibfield  {author} {\bibinfo {author} {\bibfnamefont {M.~D.}\ \bibnamefont
  {Homan}}\ and\ \bibinfo {author} {\bibfnamefont {A.}~\bibnamefont {Gelman}},\
  }\href@noop {} {\bibfield  {journal} {\bibinfo  {journal} {J. Mach. Learn.
  Res.}\ }\textbf {\bibinfo {volume} {15}},\ \bibinfo {pages} {1593–1623}
  (\bibinfo {year} {2014})}\BibitemShut {NoStop}%
\bibitem [{\citenamefont {Goodman}\ and\ \citenamefont
  {Weare}(2010)}]{goodman10}%
  \BibitemOpen
  \bibfield  {author} {\bibinfo {author} {\bibfnamefont {J.}~\bibnamefont
  {Goodman}}\ and\ \bibinfo {author} {\bibfnamefont {J.}~\bibnamefont
  {Weare}},\ }\href {\doibase 10.2140/camcos.2010.5.65} {\bibfield  {journal}
  {\bibinfo  {journal} {Comm. App. Math. Com. Sc.}\ }\textbf {\bibinfo {volume}
  {5}},\ \bibinfo {pages} {65} (\bibinfo {year} {2010})}\BibitemShut {NoStop}%
\bibitem [{\citenamefont {{Foreman-Mackey}}\ \emph {et~al.}(2013)\citenamefont
  {{Foreman-Mackey}}, \citenamefont {{Hogg}}, \citenamefont {{Lang}},\ and\
  \citenamefont {{Goodman}}}]{foremanmackey13}%
  \BibitemOpen
  \bibfield  {author} {\bibinfo {author} {\bibfnamefont {D.}~\bibnamefont
  {{Foreman-Mackey}}}, \bibinfo {author} {\bibfnamefont {D.~W.}\ \bibnamefont
  {{Hogg}}}, \bibinfo {author} {\bibfnamefont {D.}~\bibnamefont {{Lang}}}, \
  and\ \bibinfo {author} {\bibfnamefont {J.}~\bibnamefont {{Goodman}}},\ }\href
  {\doibase 10.1086/670067} {\bibfield  {journal} {\bibinfo  {journal} {PASP}\
  }\textbf {\bibinfo {volume} {125}},\ \bibinfo {pages} {306} (\bibinfo {year}
  {2013})}\BibitemShut {NoStop}%
\bibitem [{\citenamefont {Reinert}\ \emph
  {et~al.}(2018{\natexlab{a}})\citenamefont {Reinert}, \citenamefont {Krebs},\
  and\ \citenamefont {Epelbaum}}]{reinert18}%
  \BibitemOpen
  \bibfield  {author} {\bibinfo {author} {\bibfnamefont {P.}~\bibnamefont
  {Reinert}}, \bibinfo {author} {\bibfnamefont {H.}~\bibnamefont {Krebs}}, \
  and\ \bibinfo {author} {\bibfnamefont {E.}~\bibnamefont {Epelbaum}},\ }\href
  {\doibase 10.1140/epja/i2018-12516-4} {\bibfield  {journal} {\bibinfo
  {journal} {Eur. Phys. J. A}\ }\textbf {\bibinfo {volume} {54}},\ \bibinfo
  {pages} {86} (\bibinfo {year} {2018}{\natexlab{a}})}\BibitemShut {NoStop}%
\bibitem [{\citenamefont {Wesolowski}\ \emph {et~al.}()\citenamefont
  {Wesolowski}, \citenamefont {Svensson}, \citenamefont {Ekström},
  \citenamefont {Forssén}, \citenamefont {Furnstahl}, \citenamefont
  {Melendez},\ and\ \citenamefont {Phillips}}]{wesolowski21}%
  \BibitemOpen
  \bibfield  {author} {\bibinfo {author} {\bibfnamefont {S.}~\bibnamefont
  {Wesolowski}}, \bibinfo {author} {\bibfnamefont {I.}~\bibnamefont
  {Svensson}}, \bibinfo {author} {\bibfnamefont {A.}~\bibnamefont {Ekström}},
  \bibinfo {author} {\bibfnamefont {C.}~\bibnamefont {Forssén}}, \bibinfo
  {author} {\bibfnamefont {R.~J.}\ \bibnamefont {Furnstahl}}, \bibinfo {author}
  {\bibfnamefont {J.~A.}\ \bibnamefont {Melendez}}, \ and\ \bibinfo {author}
  {\bibfnamefont {D.~R.}\ \bibnamefont {Phillips}},\ }\href@noop {} {\bibfield
  {journal} {\bibinfo  {journal} {Phys. Rev. C (to be published)}\ }}\Eprint
  {http://arxiv.org/abs/2104.04441} {arXiv:2104.04441} \BibitemShut {NoStop}%
\bibitem [{\citenamefont {Sokal}(1997)}]{Sokal}%
  \BibitemOpen
  \bibfield  {author} {\bibinfo {author} {\bibfnamefont {A.}~\bibnamefont
  {Sokal}},\ }\enquote {\bibinfo {title} {Monte carlo methods in statistical
  mechanics: Foundations and new algorithms},}\ in\ \href {\doibase
  10.1007/978-1-4899-0319-8_6} {\emph {\bibinfo {booktitle} {Functional
  Integration: Basics and Applications}}},\ \bibinfo {editor} {edited by\
  \bibinfo {editor} {\bibfnamefont {C.}~\bibnamefont {DeWitt-Morette}},
  \bibinfo {editor} {\bibfnamefont {P.}~\bibnamefont {Cartier}}, \ and\
  \bibinfo {editor} {\bibfnamefont {A.}~\bibnamefont {Folacci}}}\ (\bibinfo
  {publisher} {Springer US},\ \bibinfo {address} {Boston, MA},\ \bibinfo {year}
  {1997})\ pp.\ \bibinfo {pages} {131--192}\BibitemShut {NoStop}%
\bibitem [{\citenamefont {Gelman}\ and\ \citenamefont
  {Rubin}(1992)}]{gelman92}%
  \BibitemOpen
  \bibfield  {author} {\bibinfo {author} {\bibfnamefont {A.}~\bibnamefont
  {Gelman}}\ and\ \bibinfo {author} {\bibfnamefont {D.~B.}\ \bibnamefont
  {Rubin}},\ }\href {\doibase 10.1214/ss/1177011136} {\bibfield  {journal}
  {\bibinfo  {journal} {Stat. Sci.}\ }\textbf {\bibinfo {volume} {7}},\
  \bibinfo {pages} {457 } (\bibinfo {year} {1992})}\BibitemShut {NoStop}%
\bibitem [{\citenamefont {Brooks}\ and\ \citenamefont
  {Gelman}(1998)}]{brooks98}%
  \BibitemOpen
  \bibfield  {author} {\bibinfo {author} {\bibfnamefont {S.}~\bibnamefont
  {Brooks}}\ and\ \bibinfo {author} {\bibfnamefont {A.}~\bibnamefont
  {Gelman}},\ }\href {\doibase 10.1080/10618600.1998.10474787} {\bibfield
  {journal} {\bibinfo  {journal} {J. Comput. Graphi. Stat.}\ }\textbf {\bibinfo
  {volume} {7}},\ \bibinfo {pages} {434} (\bibinfo {year} {1998})}\BibitemShut
  {NoStop}%
\bibitem [{\citenamefont {Gelman}\ \emph {et~al.}(2014)\citenamefont {Gelman},
  \citenamefont {Carlin}, \citenamefont {Stern}, \citenamefont {Rubin},
  \citenamefont {Vehtari},\ and\ \citenamefont {Dunson}}]{bda3}%
  \BibitemOpen
  \bibfield  {author} {\bibinfo {author} {\bibfnamefont {A.}~\bibnamefont
  {Gelman}}, \bibinfo {author} {\bibfnamefont {J.~B.}\ \bibnamefont {Carlin}},
  \bibinfo {author} {\bibfnamefont {H.~S.}\ \bibnamefont {Stern}}, \bibinfo
  {author} {\bibfnamefont {D.~B.}\ \bibnamefont {Rubin}}, \bibinfo {author}
  {\bibfnamefont {A.}~\bibnamefont {Vehtari}}, \ and\ \bibinfo {author}
  {\bibfnamefont {D.~B.}\ \bibnamefont {Dunson}},\ }\href@noop {} {\emph
  {\bibinfo {title} {Bayesian data analysis}}},\ Texts in statistical science
  series\ (\bibinfo  {publisher} {Chapman \& Hall/CRC},\ \bibinfo {year}
  {2014})\BibitemShut {NoStop}%
\bibitem [{\citenamefont {Vehtari}\ \emph {et~al.}(2021)\citenamefont
  {Vehtari}, \citenamefont {Gelman}, \citenamefont {Simpson}, \citenamefont
  {Carpenter},\ and\ \citenamefont {Bürkner}}]{Vehtari2021}%
  \BibitemOpen
  \bibfield  {author} {\bibinfo {author} {\bibfnamefont {A.}~\bibnamefont
  {Vehtari}}, \bibinfo {author} {\bibfnamefont {A.}~\bibnamefont {Gelman}},
  \bibinfo {author} {\bibfnamefont {D.}~\bibnamefont {Simpson}}, \bibinfo
  {author} {\bibfnamefont {B.}~\bibnamefont {Carpenter}}, \ and\ \bibinfo
  {author} {\bibfnamefont {P.-C.}\ \bibnamefont {Bürkner}},\ }\href {\doibase
  10.1214/20-BA1221} {\bibfield  {journal} {\bibinfo  {journal} {Bayesian
  Anal.}\ }\textbf {\bibinfo {volume} {16}},\ \bibinfo {pages} {667 } (\bibinfo
  {year} {2021})}\BibitemShut {NoStop}%
\bibitem [{\citenamefont {Hammersley}\ and\ \citenamefont
  {Morton}(1956)}]{hammersley_morton_1956}%
  \BibitemOpen
  \bibfield  {author} {\bibinfo {author} {\bibfnamefont {J.~M.}\ \bibnamefont
  {Hammersley}}\ and\ \bibinfo {author} {\bibfnamefont {K.~W.}\ \bibnamefont
  {Morton}},\ }\href {\doibase 10.1017/S0305004100031455} {\bibfield  {journal}
  {\bibinfo  {journal} {Math. Proc. Camb. Philos. Soc.}\ }\textbf {\bibinfo
  {volume} {52}},\ \bibinfo {pages} {449–475} (\bibinfo {year}
  {1956})}\BibitemShut {NoStop}%
\bibitem [{\citenamefont {Feroz}\ \emph {et~al.}(2009)\citenamefont {Feroz},
  \citenamefont {Hobson},\ and\ \citenamefont
  {Bridges}}]{10.1111/j.1365-2966.2009.14548.x}%
  \BibitemOpen
  \bibfield  {author} {\bibinfo {author} {\bibfnamefont {F.}~\bibnamefont
  {Feroz}}, \bibinfo {author} {\bibfnamefont {M.~P.}\ \bibnamefont {Hobson}}, \
  and\ \bibinfo {author} {\bibfnamefont {M.}~\bibnamefont {Bridges}},\ }\href
  {\doibase 10.1111/j.1365-2966.2009.14548.x} {\bibfield  {journal} {\bibinfo
  {journal} {Mon. Not. R. Astron. Soc.}\ }\textbf {\bibinfo {volume} {398}},\
  \bibinfo {pages} {1601} (\bibinfo {year} {2009})}\BibitemShut {NoStop}%
\bibitem [{\citenamefont {K\"onig}\ \emph {et~al.}(2020)\citenamefont
  {K\"onig}, \citenamefont {Ekstr\"om}, \citenamefont {Hebeler}, \citenamefont
  {Lee},\ and\ \citenamefont {Schwenk}}]{Konig:2019adq}%
  \BibitemOpen
  \bibfield  {author} {\bibinfo {author} {\bibfnamefont {S.}~\bibnamefont
  {K\"onig}}, \bibinfo {author} {\bibfnamefont {A.}~\bibnamefont {Ekstr\"om}},
  \bibinfo {author} {\bibfnamefont {K.}~\bibnamefont {Hebeler}}, \bibinfo
  {author} {\bibfnamefont {D.}~\bibnamefont {Lee}}, \ and\ \bibinfo {author}
  {\bibfnamefont {A.}~\bibnamefont {Schwenk}},\ }\href {\doibase
  10.1016/j.physletb.2020.135814} {\bibfield  {journal} {\bibinfo  {journal}
  {Phys. Lett. B}\ }\textbf {\bibinfo {volume} {810}},\ \bibinfo {pages}
  {135814} (\bibinfo {year} {2020})}\BibitemShut {NoStop}%
\bibitem [{\citenamefont {Ekstr\"om}\ and\ \citenamefont
  {Hagen}(2019)}]{Ekstrom:2019lss}%
  \BibitemOpen
  \bibfield  {author} {\bibinfo {author} {\bibfnamefont {A.}~\bibnamefont
  {Ekstr\"om}}\ and\ \bibinfo {author} {\bibfnamefont {G.}~\bibnamefont
  {Hagen}},\ }\href {\doibase 10.1103/PhysRevLett.123.252501} {\bibfield
  {journal} {\bibinfo  {journal} {Phys. Rev. Lett.}\ }\textbf {\bibinfo
  {volume} {123}},\ \bibinfo {pages} {252501} (\bibinfo {year}
  {2019})}\BibitemShut {NoStop}%
\bibitem [{\citenamefont {Melendez}\ \emph {et~al.}()\citenamefont {Melendez},
  \citenamefont {Drischler}, \citenamefont {Garcia}, \citenamefont
  {Furnstahl},\ and\ \citenamefont {Zhang}}]{melendez2021fast}%
  \BibitemOpen
  \bibfield  {author} {\bibinfo {author} {\bibfnamefont {J.~A.}\ \bibnamefont
  {Melendez}}, \bibinfo {author} {\bibfnamefont {C.}~\bibnamefont {Drischler}},
  \bibinfo {author} {\bibfnamefont {A.~J.}\ \bibnamefont {Garcia}}, \bibinfo
  {author} {\bibfnamefont {R.~J.}\ \bibnamefont {Furnstahl}}, \ and\ \bibinfo
  {author} {\bibfnamefont {X.}~\bibnamefont {Zhang}},\ }\href@noop {} {}\Eprint
  {http://arxiv.org/abs/2106.15608} {arXiv:2106.15608} \BibitemShut {NoStop}%
\bibitem [{\citenamefont {Miller}\ \emph {et~al.}()\citenamefont {Miller},
  \citenamefont {Ekström},\ and\ \citenamefont
  {Forssén}}]{miller2021accelerating}%
  \BibitemOpen
  \bibfield  {author} {\bibinfo {author} {\bibfnamefont {S.~B.~S.}\
  \bibnamefont {Miller}}, \bibinfo {author} {\bibfnamefont {A.}~\bibnamefont
  {Ekström}}, \ and\ \bibinfo {author} {\bibfnamefont {C.}~\bibnamefont
  {Forssén}},\ }\href@noop {} {}\Eprint {http://arxiv.org/abs/2106.00454}
  {arXiv:2106.00454} \BibitemShut {NoStop}%
\bibitem [{\citenamefont {von Przewoski}\ \emph {et~al.}(1998)\citenamefont
  {von Przewoski} \emph {et~al.}}]{vonPrzewoski:1998ye}%
  \BibitemOpen
  \bibfield  {author} {\bibinfo {author} {\bibfnamefont {B.}~\bibnamefont {von
  Przewoski}} \emph {et~al.},\ }\href {\doibase 10.1103/PhysRevC.58.1897}
  {\bibfield  {journal} {\bibinfo  {journal} {Phys. Rev. C}\ }\textbf {\bibinfo
  {volume} {58}},\ \bibinfo {pages} {1897} (\bibinfo {year}
  {1998})}\BibitemShut {NoStop}%
\bibitem [{sup()}]{suppl}%
  \BibitemOpen
  \href@noop {} {}\bibinfo {note} {See Supplemental Material at [URL will be
  inserted by publisher] for posterior predictive distributions at LO, NLO, and
  NNLO for the nucleon-nucleon scattering cross sections in the Granada
  database.}\BibitemShut {Stop}%
\bibitem [{\citenamefont {Hoshizaki}(1969)}]{Hoshizaki:1969qt}%
  \BibitemOpen
  \bibfield  {author} {\bibinfo {author} {\bibfnamefont {N.}~\bibnamefont
  {Hoshizaki}},\ }\href {\doibase 10.1143/PTPS.42.107} {\bibfield  {journal}
  {\bibinfo  {journal} {Prog. Theor. Phys. Suppl.}\ }\textbf {\bibinfo {volume}
  {42}},\ \bibinfo {pages} {107} (\bibinfo {year} {1969})}\BibitemShut
  {NoStop}%
\bibitem [{\citenamefont {Melendez}\ \emph {et~al.}(2017)\citenamefont
  {Melendez}, \citenamefont {Wesolowski},\ and\ \citenamefont
  {Furnstahl}}]{Melendez:2017phj}%
  \BibitemOpen
  \bibfield  {author} {\bibinfo {author} {\bibfnamefont {J.~A.}\ \bibnamefont
  {Melendez}}, \bibinfo {author} {\bibfnamefont {S.}~\bibnamefont
  {Wesolowski}}, \ and\ \bibinfo {author} {\bibfnamefont {R.~J.}\ \bibnamefont
  {Furnstahl}},\ }\href {\doibase 10.1103/PhysRevC.96.024003} {\bibfield
  {journal} {\bibinfo  {journal} {Phys. Rev. C}\ }\textbf {\bibinfo {volume}
  {96}},\ \bibinfo {pages} {024003} (\bibinfo {year} {2017})}\BibitemShut
  {NoStop}%
\bibitem [{\citenamefont {Reinert}\ \emph
  {et~al.}(2018{\natexlab{b}})\citenamefont {Reinert}, \citenamefont {Krebs},\
  and\ \citenamefont {Epelbaum}}]{Reinert:2017usi}%
  \BibitemOpen
  \bibfield  {author} {\bibinfo {author} {\bibfnamefont {P.}~\bibnamefont
  {Reinert}}, \bibinfo {author} {\bibfnamefont {H.}~\bibnamefont {Krebs}}, \
  and\ \bibinfo {author} {\bibfnamefont {E.}~\bibnamefont {Epelbaum}},\ }\href
  {\doibase 10.1140/epja/i2018-12516-4} {\bibfield  {journal} {\bibinfo
  {journal} {Eur. Phys. J. A}\ }\textbf {\bibinfo {volume} {54}},\ \bibinfo
  {pages} {86} (\bibinfo {year} {2018}{\natexlab{b}})},\ \Eprint
  {http://arxiv.org/abs/1711.08821} {arXiv:1711.08821 [nucl-th]} \BibitemShut
  {NoStop}%
\bibitem [{\citenamefont {Yang}\ \emph {et~al.}(2021)\citenamefont {Yang},
  \citenamefont {Ekstr\"om}, \citenamefont {Forss\'en},\ and\ \citenamefont
  {Hagen}}]{Yang:2020pgi}%
  \BibitemOpen
  \bibfield  {author} {\bibinfo {author} {\bibfnamefont {C.}~\bibnamefont
  {Yang}}, \bibinfo {author} {\bibfnamefont {A.}~\bibnamefont {Ekstr\"om}},
  \bibinfo {author} {\bibfnamefont {C.}~\bibnamefont {Forss\'en}}, \ and\
  \bibinfo {author} {\bibfnamefont {G.}~\bibnamefont {Hagen}},\ }\href
  {\doibase 10.1103/PhysRevC.103.054304} {\bibfield  {journal} {\bibinfo
  {journal} {Phys. Rev. C}\ }\textbf {\bibinfo {volume} {103}},\ \bibinfo
  {pages} {054304} (\bibinfo {year} {2021})}\BibitemShut {NoStop}%
\end{thebibliography}%
